\documentclass[sigconf]{acmart}
\usepackage{booktabs} 
\usepackage{bm}

\usepackage{graphicx}
\usepackage{subfig}
\usepackage{url}
\usepackage{multirow}

\usepackage{amsmath,bm,amssymb}
\usepackage{algorithm}
\usepackage{algpseudocode}
\usepackage{varwidth}

\usepackage[flushleft]{threeparttable}
\algdef{SE}[DOWHILE]{Do}{doWhile}{\algorithmicdo}[1]{\algorithmicwhile\ #1}%

\usepackage{algorithm} 
\usepackage{amsmath}
\usepackage{mathtools}
\usepackage{url}
\newsavebox{\measurebox}
\renewcommand{\footnotesize}{\small}
\usepackage{amsfonts}
\usepackage[flushleft]{threeparttable}
\usepackage{url}

\usepackage{amssymb}

\usepackage{tikz}
\usetikzlibrary{calc}

\usepackage{colortbl}
\usepackage{booktabs}
\newcolumntype{L}[1]{>{\raggedright\let\newline\\\arraybackslash\hspace{0pt}}m{#1}}
\newcolumntype{C}[1]{>{\centering\let\newline\\\arraybackslash\hspace{0pt}}m{#1}}

\newcommand{\cev}[1]{\reflectbox{\ensuremath{\vec{\reflectbox{\ensuremath{#1}}}}}}

\usepackage{flushend}

\copyrightyear{2020}
\acmYear{2020} 
\setcopyright{iw3c2w3}
\acmConference[WWW '20]{Proceedings of The Web Conference 2020}{April 20--24, 2020}{Taipei, Taiwan} 
\acmBooktitle{Proceedings of The Web Conference 2020 (WWW '20), April 20--24, 2020, Taipei, Taiwan}
\acmPrice{}
\acmDOI{10.1145/3366423.3380116}
\acmISBN{978-1-4503-7023-3/20/04}

\begin{document}
\title{Future Data Helps Training: Modeling Future Contexts for Session-based  Recommendation}

\author{Fajie Yuan}
\affiliation{%
	\institution{Tencent}
	\city{Shenzhen} 
	\state{China} 
}
\email{fajieyuan@tencent.com }

\author{Xiangnan He}
\affiliation{%
	\institution{University of Science and Technology of China}
	\city{Hefei} 
	\state{China} 
}
\email{xiangnanhe@gmail.com}

\author{Haochuan Jiang}
\affiliation{%
	\institution{University of Edinburgh}
	\city{Edinburgh} 
	\state{UK} 
}
\email{haochuan.jiang@ed.ac.uk}
\authornote{\scriptsize Work mostly done at Tencent.}

\author{Guibing Guo}
\affiliation{%
	\institution{Northeastern University}
	\city{Shenyang} 
	\state{China} 
}

\email{guogb@swc.neu.edu.cn}

\author{Jian Xiong, Zhezhao Xu, Yilin Xiong}
\affiliation{%
	\institution{Tencent}
	\city{Shenzhen} 
	\state{China} 
}
\email{{janexiong, zhezhaoxu, plutoxiong} @tencent.com }

\begin{abstract}
Session-based recommender systems have attracted much attention recently.  
To capture the sequential dependencies, 
existing methods resort either to data augmentation techniques or 
left-to-right style autoregressive training. 
Since these methods are aimed to model the sequential nature of user behaviors, they ignore the future data of a target interaction when constructing the prediction model for it. However, we argue that the future interactions after a target interaction, which are also available during training, provide valuable signal on user preference and can be used to enhance the recommendation quality.  

Properly integrating future data into model training, however, is non-trivial to achieve, since it disobeys machine learning principles and can easily cause data leakage. To this end, we propose a new encoder-decoder framework named \textit{Gap-filling based Recommender} (GRec), which trains the encoder and decoder by a gap-filling mechanism. Specifically, the encoder takes a partially-complete session sequence (where some items are masked by purpose) as input, and the decoder predicts these masked items conditioned on the encoded representation. 
We instantiate the general GRec framework using convolutional neural network with sparse kernels, giving consideration to both accuracy and efficiency. 
We conduct experiments on two real-world datasets covering short-, medium-, and long-range user sessions, showing that GRec significantly outperforms the state-of-the-art sequential recommendation methods. More empirical studies verify the high utility of modeling future contexts under our GRec framework.

\end{abstract}

%
%





\maketitle

\section{Introduction}
\label{Introduction}

Session-based Recommender system (SRS) has become an emerging topic in the recommendation domain, 
which 
aims to predict the next item based on an ordered history of 
interacted
items within a user session. While recent advances in deep neural networks
\cite{hidasi2015session,quadrana2017personalizing,Tuan:2017:CNS:3109859.3109900,tang2018caser} 
are effective in modeling
user short-term interest transition,
it remains as a fundamental challenge to 
capture the sequential dependencies in long-range sessions~\cite{yuan2019simple,tang2019towards,ma2019hierarchical}.
In practice, long-range user  sessions  widely exist in scenarios such as micro-video and news recommendations. For example, users on TikTok\footnote{\scriptsize \url{https://www.tiktok.com}} 
may watch 100
micro-videos in 30 minutes as the average playing time of each video takes only 15 seconds.



Generally speaking, there are two popular strategies 
to train recommender models from sequential data:
data augmentation \cite{tan2016improved,Tuan:2017:CNS:3109859.3109900,li2017neural,tan2016improved,tang2018caser,fang2019deep} and autoregressive training~\cite{yuan2019simple,kang2018self}. Specifically, the data augmentation approach, such as the improved GRU4Rec~\cite{tan2016improved}, performs data preprocessing and generates new training sub-sessions by using prefixes of the target sequence, and the recommender then predicts the last item in the sequence. The autoregressive approach models the  distribution of an entire sequence in an end-to-end  manner, rather than only the last item. 
This idea results in a typical left-to-right style  unidirectional generative model, referred to as NextItNet~\cite{yuan2019simple}.
The two strategies share similar intuition in that when constructing the prediction function for a target interaction, only its past user behaviors (which we also term as ``contexts'' in this paper)  are taken into account.

In standard sequential data prediction, it is a straightforward and reasonable choice to predict a target entry based on the past entries~\cite{van2016conditional,kalchbrenner2017video}. However, in sequential recommendation, we argue that such a choice may limit the model's ability. The key reason is that although user behaviors are in the form of sequence data, the sequential dependency may not be strictly held. For example, after a user purchases a phone, she may click phone case, earphone, and screen protector in the session, but there is no sequential dependency among the three items --- in other words, it is likely that the user clicks the three items in any order. As such, it is not compulsory to model a user session as a strict sequence. Secondly, the objective of recommendation is to accurately estimate a user's preference, and using more data is beneficial to the preference estimation. As the future data after a target interaction also evidences the user's preference, it is reasonable to believe that modeling the future data can help build better prediction model for the target interaction.

\begin{figure}[t]
	\centering
	\subfloat[\scriptsize Typical seq2seq learning ($\surd$). ]{\includegraphics[width=0.25\textwidth]{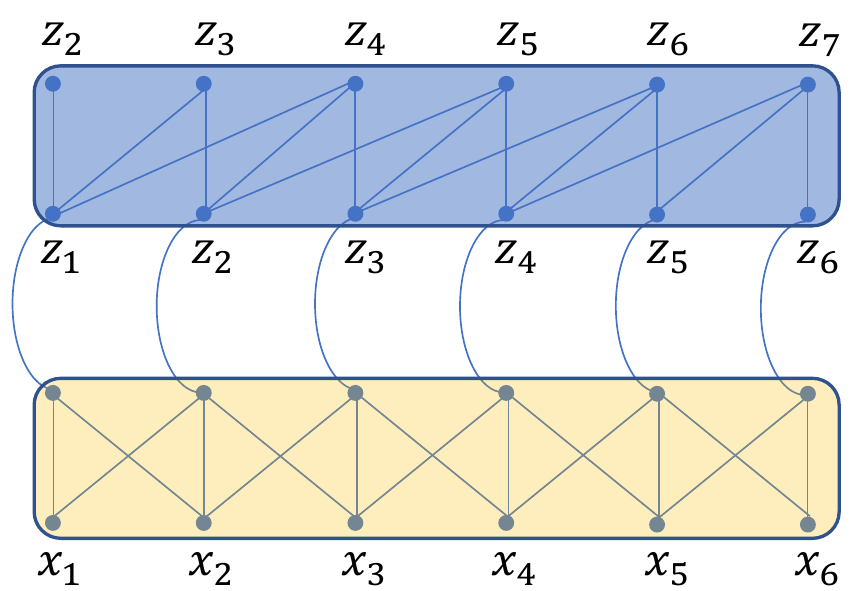}\label{vanallia}}
	\subfloat[\scriptsize Seq2seq learning for SRS ($\times$). ]{\includegraphics[width=0.25\textwidth]{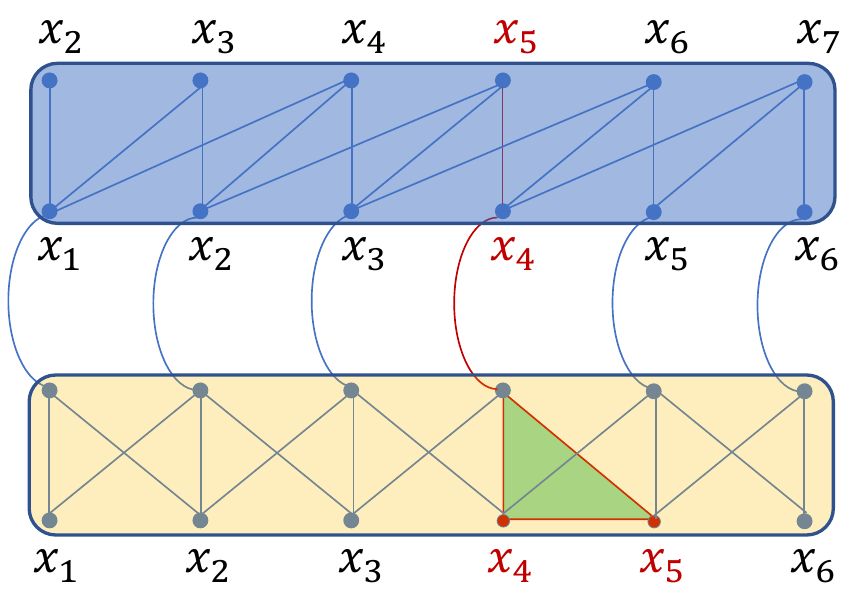}\label{dilated}}
	\caption{\small Examples of ED architecture to model sequential data (encoder as yellow and decoder as blue). 
		(a) is a standard ED architecture where the input $x$ and output $z$ are from two different domains. E.g., in English-to-Chinese machine translation, $x$ and $z$ represent English and Chinese words respectively. 
		(b) is a direct application of ED on SRS with future data modeled. As the predicted items (e.g., $x_5$ with red color) by the decoder can be observed from the encoder's input, it causes data leakage in training.  
	}
	\label{standardEDSRS}
\end{figure}

Nevertheless, it is challenging to model with the future data well, since it disobeys machine learning principles and can cause data leakage if not handled properly. Taking the encoder-decoder (ED) neural architecture as an example, which has been extensively used in sequential data modeling~\cite{sutskever2014sequence,bahdanau2014neural, kalchbrenner2016neural}. 
As illustrated in Figure~\ref{standardEDSRS} (a), in machine translation, when predicting a target word in a sequence (i.e., sentence), the encoder takes the words from both sides as the input source. Since the source and target words are from different domains, there is no issue of data leakage. However, if we apply the same ED architecture to user session modeling, as illustrated in Figure~\ref{standardEDSRS} (b), the data leakage issue arises inevitably. This is because the source and target entries are from the same domain, such that a target entry (to be predicted by the decoder) exactly occurs in the input of the encoder.  


To address the above issues, 
we propose a new SRS method that models the future contexts:
\textit{Gap-filling based encoder-decoder framework for sequential Recommendation}, or GRec for short. 
GRec revises the ED design by tailoring it for future data modeling without data leakage:
the encoder and decoder are jointly trained by a gap-filling mechanism~\cite{sakaguchi2013discriminative}, which is inspired by the recent development of pretrained language model~\cite{devlin2018bert}. 
Specifically,  a portion of items in a user session are deleted by filling in the gap symbols (e.g., "$\_\_$"). 
The encoder takes the partially-complete sequence as the input, and the decoder predicts the items of these gaps conditioned on the encoded representation model. Through this way, GRec can force the encoder to be aware of the general user preference, represented by unmasked actions, and simultaneously force the decoder to perform next item generation conditioned on  both the past contexts
and the encoded general user preference. 

The contributions of the work are listed as follows:
\begin{itemize}
	\item We highlight the necessity of modeling future contexts in session-based recommender system, and develop a general neural network framework GRec that works without data leakage. 
	\item We specify GRec using convolutional neural network with sparse kernels~\cite{yuan2019simple}, unifying the advantages of both autoregressive mechanism for sequence generation and two-side contexts for encoding.
	\item We propose a projector neural network with an inverted bottleneck architecture in the decoder, which can enhance the representational bandwidth between the encoder and the decoder.
	\item We conduct extensive experiments on two real-world datasets, justifying the effectiveness of GRec in leveraging future contexts for session-based recommender system. 
\end{itemize}

The paper is organized as follows. In Section 2, we review recent advancements in using sequential neural network models for SRS. Particularly, we recapitulate two widely used unidirectional training approaches. 
In Section 3, we first investigate the straight ways to model  bidirectional contexts within a user session, and point out the drawbacks of them for the item recommendation task. After that, we describe in detail the  framework and architecture of our proposed GRec. In Section 4, we conduct experiments and ablation tests to verfiy the effectiveness of  GRec in the SRS task.  In Section 5, we draw conclusions and future work.


\section{Preliminaries}
In this section, we first define the problem of session-based recommendations. Then,  we recapitulate two state-of-the-art left-to-right style sequential recommendation methods.   At last, we review previous work of SRS. 
\subsection{Top-$N$ Session-based Recommendation}
 
The formulation of top-$N$ session-based recomendation in this paper closely follows that in  \cite{tang2018caser,tan2016improved,yuan2019simple}. In SRS, the concept ``session" is defined as a collection of  items (referring to any objects e.g., videos, songs or queries) that happened at one time or in a certain period of time  \cite{li2017neural,wang2019survey}. For instance, both a list of browsed webpages and a collection of watched videos consumed in an hour or a day  can be regarded as a session. Formally, 
let $\{x_1,...,x_{t-1},x_t\}$ be a user session with items in the chronological order, where $x_i \in \mathbb{R}^n$ $(1\leq  i \leq t)$ denotes the index of a clicked item out of a total number of $n$ items in the session.  
The task of SRS is to train a  model so that for a given prefix session data, $x=\{x_1,...,x_{i}\}$, it can generate the distribution $\bm{\mathit{\hat{y}}}$ for items which will occur in the future,
where $\bm{\mathit{\hat{y}}} =[ \hat{y}_1,...,\hat{y}_n ] \in \mathbb{R}^n $. $\hat{y}_j$  represents probablity value of item $i+1$ occurring in the next clicking event.
 In practice, SRS typically makes more than one recommendation by selecting the top-$N$ (e.g., $N=10$) items from $\bm{\mathit{\hat{y}}}$, referred to as the top-$N$ session-based recommendations.

\subsection{The Left-to-Right-style Algorithms}
\label{issuesofnextitnet}
In this section, we mainly review the sequential recommendation models that have the left-to-right fashions, including but not limited to Improved GRU4Rec~\cite{tan2016improved} (short for IGRU4Rec), Caser~\cite{tang2018caser}, and NextItNet~\cite{yuan2019simple}. 
Among these models, IGRU4Rec and Caser fall in the line of data augmentation methods, as shown Figure~\ref{dataaugmentation}~(a), while NextItNet is a typical AR-based generative model, as shown in  Figure~\ref{dataaugmentation}~(b). Note, GRU4Rec, NextItNet can be trained by both DA and AR methods.
\subsubsection{\textbf{Data Augmentation }}
\label{DA}
\begin{figure}[!t]
	\centering
	\subfloat[\scriptsize  Data augmentation (DA). ]{\includegraphics[width=0.24\textwidth]{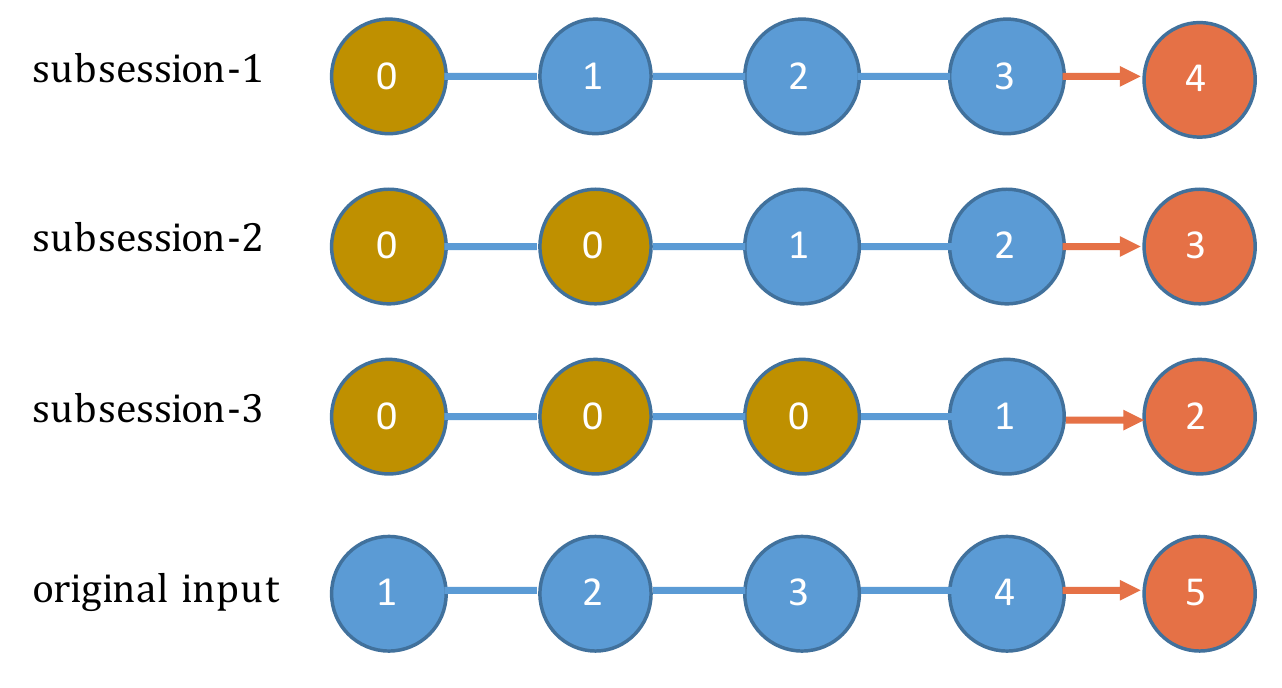}\label{vanallia}}
	\subfloat[\scriptsize Autoregressive models (AR). ]{\includegraphics[width=0.245\textwidth]{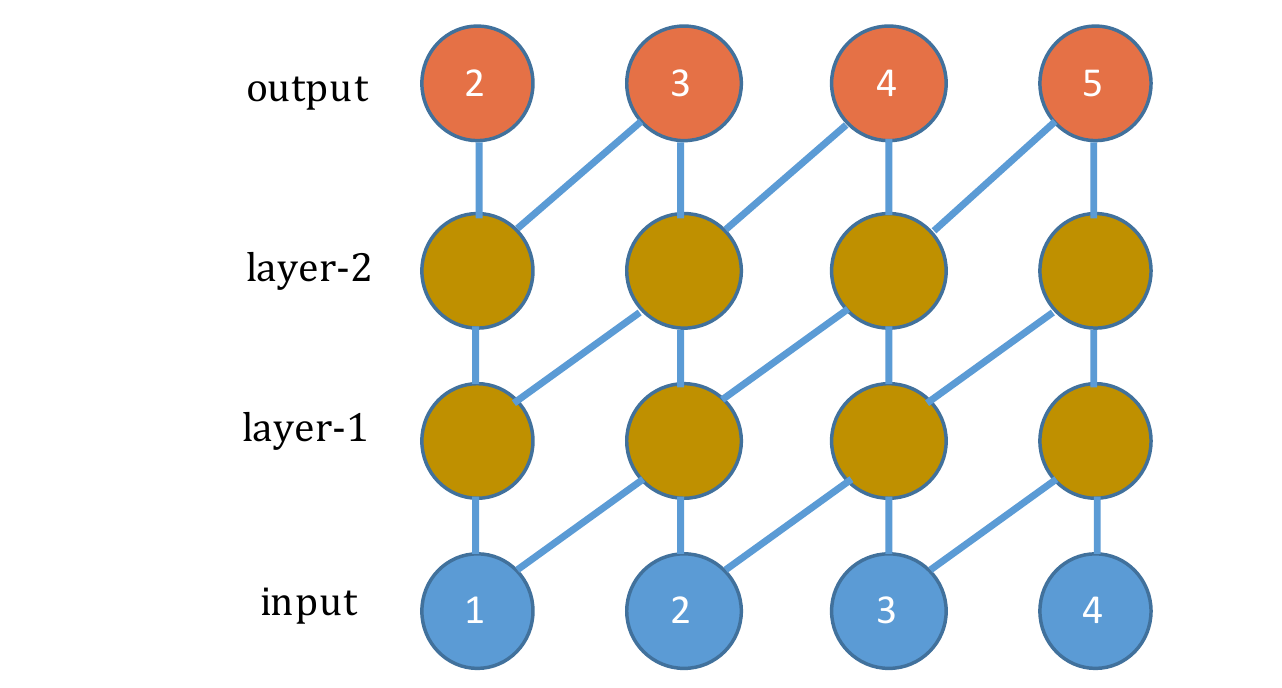}\label{dilated}}
	\caption{\small 
		Two techniques to train sequential recommendation models. The  numbers represent observed itemIDs in each user session. "0" is the padding token. The red token represents
		the items to be predicted by SRS.
		\text{(a)} The typical data augmentation approach  with a number of new  subsessions created by spliting the original input session. 
		\text{(b)}The  typical left-to-right style autoregressive approach.
		The  item  that is being predicted  is only determined by its previous timesteps, i.e., $p(x_t)=p(x_t|x_1,...,x_{t-1})$. For instance,
		item ``4'' is predicted by ``1, 2, 3'' which achieves the same effect with session-1 in (a).
		The overall training objectives in (b) can be regarded as the sum of the separate objective of all subsessions in (a). 
	}
	\label{dataaugmentation}
\end{figure}

The authors in \cite{tan2016improved} proposed a generic data augmentation method to improve recommendation quality of SRS, which has been further applied in a majority of future work, such as \cite{Tuan:2017:CNS:3109859.3109900,li2017neural,tan2016improved,tang2018caser}. The basic idea of DA in SRS is to treat all prefixes in the user session as new training sequences~\cite{hidasi2015session}.
Specifically, for a given user session   $\{x_1,...,x_{t-1},x_t\}$,  the DA method will generate 
a collection of sequences and target labels \{$(x_2|x_1)$, $(x_3|x_1,x_2)$,..., $(x_{t}|x_1,x_2,...,x_{t-1})$\} as illustrated in Figure~\ref{dataaugmentation} (a).
Following this processing, the sequential model is able to learn all conditional dependencies rather than only the last item $x_{t}$ and the prefix sequence  $\{x_1,x_2,...,x_{t-1}\}$. Due to more information learned by additional subsessions, data augmentation becomes an effective way to reduce the overfitting problem especially when the user session is longer and the user-item matrix is sparse. Even though the  data augmentation method has been successfully applied in
numerous SRS work, it may lead to a break regarding the integrity of the entire user session and significantly increase training times~\cite{yuan2019simple}. 

\subsubsection{\textbf{Autoregressive Models (AR)}}
The AR-style learning methods~\cite{yuan2019simple,kang2018self} propose to  optimizing all positions of the original input sequence rather than only the final one. 
 Specifically, the generative model takes $\{x_1,...,x_{t-1}\}$ (or $x_{1:{t-1}}$)
as the input and output probabilities (i.e., softmax) over 
$x_{2:{t}}$ by a seq2seq (sequence-to-sequence) manner. Mathematically, the joint distribution of a user session $\{x_1,...,x_{t-1},x_t\}$  can be factorized out as a product of conditional distributions following the chain rule:
\begin{equation}
\label{jointdis}
p(x)=\prod_{i=1}^{t}p(x_i|x_1,...,x_{i-1}; {\Theta})
\end{equation}
where  $p(x_i|x_1,...,x_{i-1})$ denotes the probability of $i$-th item $x_i$ conditioned on its all prefix $x_{1:{i-1}}$, ${\Theta}$ is the parameters. With this formulation, each predicted item can be conditioned on all items that are clicked earlier. Correspondingly, the AR method does not rely on the data augmentation technique any more.

%

As mentioned, both the data augmentation and AR  approaches train the user session  in an order from left to right.
Though it conforms to the generation law of sequential data with natural orders, the way of modeling inevitably neglects many useful future contexts that associate with the target interaction. Particularly in the field of recommendation, user behaviors  in the sequence may not obey rigid order relations.
Hence, these methods  may limit the ability of sequential recommendation models. Moreover,
leveraging the addittional future contexts can also be regarded as a way of data augmentation that helps models alleviate the sparsity problem in SRS.
 Motivated by this,  we believe that it is crucial to investigate the impact to sequential recomendation models by taking into account both directional contexts.

\subsection{Related Work}

Recently, the powerful deep neural network based sequential models  have almost dominated the field of session-based recommender systems (SRS). Among these models, GRU4Rec \cite{hidasi2015session}  is  regarded as the pioneering work that employs the recurrent neural network (RNN) to model the evolution of user preference. Inspired by the success, a class of RNN-based models  has been developed. For example, an improved RNN variant in \cite{tan2016improved} showed promising improvements over standard RNN models by proposing data augmentation techniques. Hidasi et al \cite{hidasi2017recurrent} further  proposed a family of alternative ranking objective functions along with effective sampling tricks to improve the cross-entropy and pairwise losses.  \cite{quadrana2017personalizing} proposed personalized SRS, while ~\cite{gu2016learning, Elena} explored how to use content and context features to enhance the recommendation accuracy. 

Another line of research work is based on convolutional neural networks (CNN) and attention mechanisms. The main reason is that RNN-based sequential models seriously depend on a hidden state from all the past interactions that cannot fully utilize parallel processing power of GPUs~\cite{yuan2019simple}. As a result, their speeds are limited in both training and evaluation. Instead, CNN and purely attention based models are inherently easier to be parallelized since all timesteps in the user session are known during training. The most typical CNN models for SRS is Caser~\cite{tang2018caser}, which treats the item embedding matrix as an image and then performs 2D convolution on it.  In NextItNet~\cite{yuan2019simple}, authors argued that  the standard CNN architecture and max pooling operation of Caser were not well-suited to model long-range user sequence. Correspondingly, they proposed using stacked dilated CNN to increase the receptive field of higher layer neurons. Moreover, authors claimed that the data augmentation techniques widely used previous work could be simply omitted by developing a seq2seq style objective function. They showed that the autoregressive NextItNet is more powerful than Caser and more efficient than RNN models for top-$N$ session-based recommendation task. Inspired by the success of Caser and NextItNet, several extended work, e.g., ~\cite{yan2019cosrec,you2019hierarchical}, were proposed by employing (or improving) the 1D dilated CNN or 2D CNN to model user-item interaction sequence.
Meanwhile,  transformer-based self-attention~\cite{kang2018self,zhang2018next,sun2019bert4rec} models also demonstrated promising results in the area of SRS. However, it is known that the  self-attention mechanism is  computationally
more expensive than the stacked dilated CNN structure since calculating self-attention of all timesteps  requires quadratic complexity. More recently, \cite{ma2019hierarchical,tang2019towards} introduced gating networks to improve  SRS by capturing both short- and long-term sequential patterns.

The above mentioned sequential recommenders are built on 
either an encoder or a decoder architecture. 
Jointly training an encoder and decoder to model two directional contexts as well as maintain the autoregressive generative mechanism has not been explored in the existing recommendation literature.
A relatively relevant work to this paper is NARM~\cite{li2017neural}, which proposed an attention-based  `ED mechanism' for SRS. However, NARM is, in fact, a sequence-to-one architecture rather than the typical seq2seq manner in its decoder network. In other words, NARM decodes the distribution only for the final item, whereas the standard ED model decodes distributions of a complete sequence. 
By contrast, our proposed GRec is  a pseq2pseq (partial-sequence-to-partial-sequence) ED paradigm where its encoder \& decoder focus on encoding and decoding incomplete  sequences. With the design, GRec  combines the advantages of both autoregressive mechanism for sequence generation and two side contexts for encoding.

\section{Methodologies}
Before introducing the final solution, we first need to investigate some conventional ways  to incorporate future contexts. Then, we shed light on the potential drawbacks of these  methods when applying them for the generating task. 
Motivated by the analysis, we present the gap-filling (or fill-in-the-blank) based encoder-decoder generative framework, namely, GRec. In the following, we instantiate 
the proposed methods using the dilated convolutional neural network used in NextItNet, giving consideration to both accuracy and efficiency.
\subsection{Two-way Data Augmentation}
\label{2Ddataaugmentation}

A straightforward approach to take advantage of future data is to reverse the original user input sequence and train the recommendation model by feeding it both the input and reversed output. This type of two-way data augmentation approach has been effectively verified in several NLP tasks \cite{sutskever2014sequence}.
The recommendation models based on both data augmentation and AR methods can be directly applied without any modification. For instance, we show this method by using NextItNet (denoted by NextItNet+), as illustrated below.  
 
\begin{equation}
\label{lefttoright}
\begin{aligned}
  NextItNet+: &\underbrace{ \{x_1,...,x_{t-1}\}}_{input}\Rightarrow  \underbrace{\{x_2,...,x_{t}\}}_{output}\\
&\underbrace{ \{x_t,...,x_{2}\}}_{input}\Rightarrow  \underbrace{\{x_{t-1},...,x_1\}}_{output}
\end{aligned}
\end{equation}


\noindent\textbf{Issues:} 
The above two-way data augmentation may have two potential drawbacks if using for the item generating task:  (1) the left  and right contexts of  item $x_{i}$ are modeled by the same set of  parameters   or same convolutional kernels of NextitNet. While in practice the impact of the left and right contexts to $x_{i}$ can be very different. That is, the same  parameter representation is not accurate from this perspective. (2) The separate training process of the left and right contexts easily results in suboptimal performance since the parameters learned for the left contexts may be largely modified when the model trains the right contexts. In view of this, a better solution is that (1) a single optimization objective consists of both the left and right contexts simultaneously, and (2) the left and right contexts are represented by different sets of model parameters.

\subsection{Two-way NextItNets (tNextItNets)}
\label{conditionmodel}

Here, we introduce two-way NextItNets that model the past contexts in the forward direction and model the future contexts in the backward direction. Similar to  the forward NextItNet, the backward NextItNet runs over a user session in reverse, predicting the previous item conditioned on the future contexts. The claim here is  different from  \cite{yuan2019simple}, where
 both the predicted items and its future contexts require to be masked. 
 we only guarantee that the item being predicted will not be accessed by higher-layer neurons.
The formulation of backward NextItNet is $p(x)=\prod_{i=1}^{t}p(x_i|x_t,.x_{t-1},..,x_{i+1}; \cev{\Theta})$.

Both the forward and backward NextItNets will produce a hidden matrix for a user session in each convolutional layer. Let $\vec{\bm{\mathit{h}}}_{x_i}$ and $\cev{\bm{\mathit{h}}}_{x_i}$ be the item
hidden vector $x_i$ calculated by the top layer NexitItNet from the forward and backward directions respectively. To form the two-way NextItNets, we concatenate $\vec{\bm{\mathit{h}}}_{x_i}$ and $\cev{\bm{\mathit{h}}}_{x_i}$, i.e., $\bm{\mathit{h}}_{x_i}=[\vec{\bm{\mathit{h}}}_{x_i}; \cev{\bm{\mathit{h}}}_{x_i}]$.
 To combine both directions in the  objective function, we  maximize the joint log likelihood of both directions.
 
\begin{equation}
\label{newloss}
\begin{aligned}
p(x)=&\prod_{i=1}^{t}p(x_i|x_1,.x_2,..,x_{i};\Theta_e,\vec{\Theta}_{NextItNet},\Theta_s) \\ &p(x_i|x_t,.x_{t-1},..,x_{i+1}; \Theta_e,\cev{\Theta}_{NextItNet},\Theta_s)
\end{aligned}
\end{equation}

The parameters $\Theta$ consist of four parts: the bottom layer item embedding $\Theta_e$,  convolutional kernels of NextItNet $ \vec{\Theta}_{NextItNet}$ \& $ \cev{\Theta}_{NextItNet}$ and weights of softmax layer $\Theta_s$.  
The idea  here has similar spirit with the recent deep contextualized word representation (ELMo) model \cite{peters2018deep} with the exception that
 ELMo was  designed for word understanding or feature extraction tasks via a Bi-RNN encoder, while  we apply the  two-way NextItNets to solve the generating task.

\noindent\textbf{Issues:} Though tNextItNets can address the training issues mentioned in Section~\ref{2Ddataaugmentation},
 the future contexts are actually unapproachable during the generating phase.
That is, the backward NextItNet is useless when it is
 used for inference.  
The discrepancies between training and predicting may seriously hurt the final recommendation performance since the optimal parameters learned for the two-way NextItNets may be largely suboptimal for the unidirectional NextItNet.
 Another downside is that two-way NextItNets are essentially a shallow concatenation of independently trained  left-to-right and right-to-left models, which have limited expresiveness in modeling complex contextual representations.
So, it is unknown whether the proposed two-way NextItNets  perform better or not than
NextItNet, even though it utilize more contexts.

\begin{figure}[t]
	\centering
	\small	
	\includegraphics[width=0.48\textwidth]{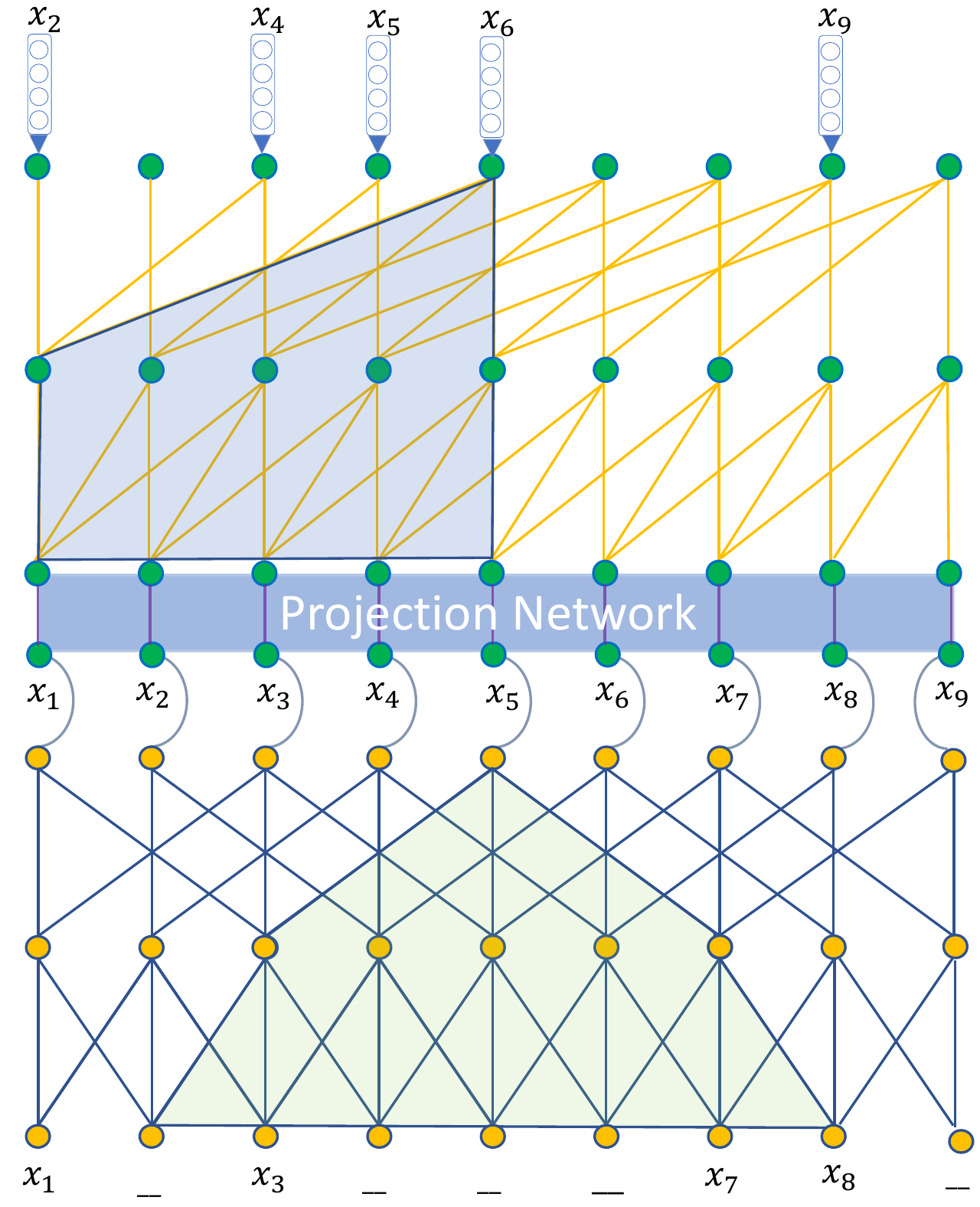}	
	\caption{\small The graphical illustration of GRec with two convolutional layers. The decoder (grean neurons) is stacked on top of the encoder (yellow neurons). The light blue \& green areas are the receptive field of $x_6$. Note the first position is not considered for masking.}
	\label{GRec}
\end{figure}
\subsection{Gap-filling Based ED framework}

In this subsection, we first present the general framework and neural architecture of GRec. Then, we discuss the relation between GRec and other popular sequential  models.
\subsubsection{ Seq2seq for SRS}

First, we introduce the basic concepts of the seq2seq learning for SRS. We denote $(x,z)\in  (\mathcal{X},\mathcal{Z})$ as a sequene pair, where $x =\{x_1,...,x_{t}\} \in \mathcal{X}$ represents the user input session sequence with $t$ items, and  $z =\{z_2,...,z_{g+1}\} \in \mathcal{Z}$ represents the output sequence, and $(\mathcal{X},\mathcal{Z})$ are regarded as source and target domains.
Unlike the standard seq2seq scenario  (i.e., Figure~\ref{standardEDSRS} (a)),  we have the following special relations in the SRS task (see Figure~\ref{standardEDSRS}~(b)):
(1) $g=t$; (2) $\{z_1,...,z_{g}\} $=$\{x_1,...,x_{t}\} $. The goal of a seq2seq model is to learn a set of  parameters $\Theta$ to describe the conditional probablity $P (z|x, \Theta)$, and usually employs the log likelihood  as the objective function \cite{sutskever2014sequence,song2019mass}:
$G(\mathcal{X},\mathcal{Z}; \Theta)=\sum_{(x,z)\in (\mathcal X,\mathcal Z)}\log p(z|x; \Theta)$. Following the decomposition of the chain rule, the probability can be further expressed as an autoregressive manner: 

\begin{equation}
\label{typicalgfedloss}
P (z|x, \Theta)=\prod_{i=2}^{g}P(z_i|z_{1:i-1},x;\Theta)=\prod_{i=2}^{t}P(x_i|x_{1:i-1},x;\Theta)
\end{equation}

\subsubsection{General Framework of Pseq2pseq} 

As can be seen, it is non-trivial to design a seq2seq learning model using Eq.~(\ref{typicalgfedloss})
since the item that is  being predicted, e.g., $x_i$, could be indirectly seen from the encoder network by $x$. 
To address this issue, 
we present the  masked-convolution operations by applying the idea of gap-filling (originally designed for the language~\cite{sakaguchi2013discriminative} task) in the ED architecture.
Here, we assume items in a user session as words in a sentence.
Correspondingly,  we could   randomly replace some  tokens in the sequence with the gap symbol ``\_\_''. The goal of gap-filling is to predict the truth of these missing tokens.

GRec consists of a modified version of encoder \& decoder, and a projector module which is injected into the decoder network. Both the encoder and decoder are described  by using the dilated convolutional neural network, although they can be simply replaced with the recurrent \cite{hidasi2015session} and attention  \cite{kang2018self} networks. The main difference of the encoder and decoder is that the encoder network is built upon the deep bidirectional CNN, while the decoder network is built upon the deep causal CNN. 
To enhance the brandwidth between the encoder and decoder, 
we  place the decoder on top of the represenation computed by the encoder,  and inject a  projector network between them.
This is in contrast to models that compress the encoder representation to a fixed-length vector~\cite{sutskever2014sequence} or align them by attention mechanism\footnote{\scriptsize We did not find the basic attention mechanisms introduced in \cite{bahdanau2014neural,vaswani2017attention} help GRec yield any better results.}~\cite{bahdanau2014neural,song2019mass}.


Formally, given a user session sequence $x =\{x_1,...,x_{t}\} \in \mathcal{X}$, we denote $\tilde{x}$ as a partial $x$, where portions of the items, i.e., $x_\triangle=\{x_{\triangle_1 },...,x_{\triangle_m }\}$ ($1 \leq m < t$), are randomly replaced with  blank mask symbols (``\_\_").   GRec optimizes a  pseq2pseq model by predicting $x_\triangle$  in each user session, taking the modified item sequence $\tilde{x}$ as input sequence. The objective function $G(\mathcal{X};\Theta)$ of GRec is defined as

\begin{equation}
\label{gfedloss}
\begin{aligned}
G(\mathcal{X};\Theta)=&\sum_{x\in \mathcal X}\log p(x_{\triangle }|\tilde{x}; \Theta)\\
=&\sum_{x\in \mathcal X}\log \prod_{i=1}^{m} p(x_{\triangle_i}|x_{1:\triangle_{i-1}},\tilde{x}; \Theta)
\end{aligned}
\end{equation}
where $\Theta$ consists of the item embeddings of encoder $\Theta_{en}$ and decoder $\Theta_{de}$,
the convolution weights of encoder $\Theta_{cnn}$ and decoder  $\vec{\Theta}_{cnn}$, the  weights of the projector  module $\Theta_{p}$ and softmax layer  $\Theta_s$. One may find that there is overlapped data between $\tilde{x}$ and $x_{1:\triangle_{i-1}}$. In fact,  since the item embeddings of encoder and decoder are not shared, the overlapped tokens in the encoder and decoder can represent different meanings.

We show the graphical example  of Eq.~(\ref{gfedloss}) using Figure \ref{GRec}.  The decoder of GRec will predict items (i.e., $x_\triangle$)  that are masked in the encoder part. As shown in Figure~\ref{GRec},	GRec takes an input sequence ``$x_1, x_3, x_7, x_8$'' and produces “$x_2, x_4, x_5, x_6, x_9$” as the output sequence. Taking the generation of item ``$x_6$'' as an example, when it is predicted, GRec  can
leverage the causal relations of the partial sequence ``$x_1, x_2, x_3, x_4, x_5$'', and meanwhile  leverage the representations of item ``$x_3, x_7, x_8$'' via the encoder, where `` $x_7, x_8$'' are the future contexts of ``$x_6$''. For clarity, we show the comparison of NextItNet (seq2seq)  and GRec (pseq2pseq) in terms of model generation as below:
 \begin{equation}
\label{compare}
\begin{aligned}
&NextItNet: \underbrace{\{x_1,x_2,x_3,...,x_7,x_8\}}_{decoder \ input}\Rightarrow \underbrace{\{x_2,x_3,x_4,...,x_8,x_9\}}_{decoder \ output}\\
&GRec: \underbrace{ \{x_1, \_\_, x_3, \_\_, \_\_, \_\_, x_7, x_8, \_\_,\}}_{encoder \ input}+\underbrace{ \{x_1,x_2,x_3,...,x_9\}}_{decoder \ input} \\
&\Rightarrow  \underbrace{\{x_2,x_4,x_5,x_6,x_9\}}_{decoder \  output}
\end{aligned}
\end{equation}
With this design, GRec can take advantage of both the past and future contexts without causing data leakage.

\subsubsection{ GRec Architecture}

In the following, we describe the components of GRec: the embedding layers,  the encoder, the decoder, the projector and the softmax layer.

\noindent \textbf{Embedding Layers.} The proposed GRec has two distinct embedding layers, namely, the encoder embedding matrix $\bm{\mathit{\widetilde{E}}} \in \mathbb{R}^{n\times d}$ and decoder embedding matrix $\bm{\mathit{\widehat{E}}}\in \mathbb{R}^{(n-1)\times d}$, where $n-1$ is the number of items and $d$ is the embedding dimension. Specifically,
the  encoder of GRec embeds the masked user input sequence $\tilde{x}$ via a look-up table from $\bm{\mathit{\widetilde{E}}} $, denoted by $\bm{\mathit{\widetilde{E}}}^{\tilde{x}} \in \mathbb{R}^{t\times d}$, while the decoder embeds the original  input sequence $x$ from $\bm{\mathit{\widehat{E}}}$, denoted by $\bm{\mathit{\widehat{E}}}^x\in \mathbb{R}^{t\times d}$. 
After the embedding  look-up operation, we denote the embeddings of the encoder and decoder as below:

 \begin{equation}
\label{emb}
\begin{aligned}
\bm{\mathit{\widetilde{E}}}_L^{\tilde{x}} =\begin{bmatrix} \bm{\mathit{\widetilde{E}}}_L^{\tilde{x}_1} & \bm{\mathit{\widetilde{E}}}_L^{\tilde{x}_0} &  \bm{\mathit{\widetilde{E}}}_L^{\tilde{x}_3} &  \bm{\mathit{\widetilde{E}}}_L^{\tilde{x}_0} &  \cdots& \bm{\mathit{\widetilde{E}}}_L^{\tilde{x}_0} \end{bmatrix}\\
\bm{\mathit{\widehat{E}}}_L^{{x}} =\begin{bmatrix} \bm{\mathit{\widehat{E}}}_L^{{x}_1}  & \bm{\mathit{\widehat{E}}}_L^{{x}_2} &\bm{\mathit{\widehat{E}}}_L^{{x}_3} & \bm{\mathit{\widehat{E}}}_L^{{x}_4} &   \cdots &\bm{\mathit{\widehat{E}}}_L^{{x}_t} \end{bmatrix}
\end{aligned}
\end{equation}
where $L$ represents the $L$-th user sequence, and $\bm{\mathit{\widetilde{E}}}_e^{\tilde{x}_0}$ represents 
the embedding vector of blank symbol, i.e., `$\_\_$', in the encoder embedding.

\noindent \textbf{Encoder:  Deep Bidirectional CNNs by Gap-filling.}
We implement the encoder network with  a series of stacked 1D dilated convolutional layers inspired by NextItNet.  To alleviate gradient vanishing issues, we wrap every two dilated layers by a residual block. 
Unlike NextItNet, the convolutional operations of the encoder are not causal. Each higher-layer neurons  can see both its left and right contexts.
With the gap-filling design, these neurons are forced to understand the unmasked contexts in the sequence. It is also worth mentioning that the proposed gap-filling mechanism is dynamic and random,  which masks different portions of the item sequence in different training batches. 
 Formally, we define the output of the encoder network  with two stacked layers    in Figure~\ref{GRec} as:
 \begin{equation}
\label{encodernet}
\begin{aligned}
\mathcal{F}_{encoder}(\bm{\mathit{\widetilde{E}}}_L^{\tilde{x}})=\bm{\mathit{\widetilde{E}}}_L^{\tilde{x}}+\mathcal{F}_{non\_cauCNN}(\bm{\mathit{\widetilde{E}}}_L^{\tilde{x}})
\end{aligned}
\end{equation}
where $\mathcal{F}_{non\_cauCNN}(\bm{\mathit{\widetilde{E}}}_L^{\tilde{x}})$ denotes the block function of non-causal  CNNs defined as

 \begin{equation}
\label{nocasuality}
\begin{aligned}
\mathcal{F}_{non\_cauCNN}(\bm{\mathit{\widetilde{E}}}_L^{\tilde{x}})=RELU(\mathcal L_n(\psi_2 (RELU(\mathcal L_n(\psi_1 (\bm{\mathit{\widetilde{E}}}_L^{\tilde{x}}))))))
\end{aligned}
\end{equation}
where $RELU$ and $\mathcal L_n$ denote non-linear activation function~\cite{nair2010rectified} and layer-normalization~\cite{ba2016layer}, $\psi_1$ and  $\psi_2$ are non-causal CNNs with 1-dilated and   2-dilated
filters respectively. 
 In practice, one can  repeat the basic encoder structure several times to capture  long-term and complex dependencies.

\noindent \textbf{Decoder: Deep  Causal CNNs by Gap-predicting.}
The decoder is composed of the embedding layer $\bm{\mathit{\widehat{E}}}_L^{{x}}$, the projector and the causal CNN modules by which each position can only attend leftward. The CNN component strictly follows NextItNet with the exception that it  is allowed to  estimate the probabilities of only the masked items in the encoder,  rather than the entire sequence in NextItNet. Meanwhile, before performing the causal CNN operations, we need to aggregate the final ouput  matrix of the encoder and the embedding matrix of
 the decoder, and then pass them into the  projector network,
which is described later. Formally, the final hidden layer (before softmax layer) of the decoder can be represented as 
 \begin{equation}
\label{decodernet}
\begin{aligned}
\mathcal{F}_{decoder}(\bm{\mathit{\widetilde{E}}}_L^{\tilde{x}},\bm{\mathit{\widehat{E}}}_L^{{x}} )=\mathcal{F}_{PR}(\bm{\mathit{\widetilde{E}}}_L^{\tilde{x}},\bm{\mathit{\widehat{E}}}_L^{{x}}) +\mathcal{F}_{cauCNN}(\mathcal{F}_{PR}(\bm{\mathit{\widetilde{E}}}_L^{\tilde{x}},\bm{\mathit{\widehat{E}}}_L^{{x}}))
\end{aligned}
\end{equation}
where 
 \begin{equation}
\label{casuality}
\begin{aligned}
&\mathcal{F}_{cauCNN}(\mathcal{F}_{PR}(\bm{\mathit{\widetilde{E}}}_L^{\tilde{x}},\bm{\mathit{\widehat{E}}}_L^{{x}}))\\
&=RELU(\mathcal L_n(\phi_2 (RELU(\mathcal L_n(\phi_1 (\mathcal{F}_{PR}(\bm{\mathit{\widetilde{E}}}_L^{\tilde{x}},\bm{\mathit{\widehat{E}}}_L^{{x}})))))))
\end{aligned}
\end{equation}
where $\mathcal{F}_{PR}(\bm{\mathit{\widetilde{E}}}_L^{\tilde{x}},\bm{\mathit{\widehat{E}}}_L^{{x}})$
and $\mathcal{F}_{cauCNN}(\mathcal{F}_{PR}(\bm{\mathit{\widetilde{E}}}_L^{\tilde{x}},\bm{\mathit{\widehat{E}}}_L^{{x}}))$ are the outputs of projection layers and causal CNN layer respectively, $\phi_1$ and  $\phi_2$ are causal CNNs with 1-dilated and 2-dilated
filters respectively.

\noindent \textbf{Projector: Connecting Encoder \& Decoder.}
Although the output hidden layer of the encoder, i.e., $\mathcal{F}_{encoder}(\bm{\mathit{\widetilde{E}}}_L^{\tilde{x}})$ and the input embedding layer of the decoder, i.e., $\bm{\mathit{\widetilde{E}}}_L^{\tilde{x}_0} $ have the same tensor shape,
we empirically find that directly placing the decoder on top of the encoder by element-wise addition may not  offer the best results. To maximize the representational brandwith between the encoder and decoder, we propose an additional projection  network  (or projector in short) in the  decoder. Specifically, the projector is an inverted bottleneck residual architecture, which  
consists of  the projection-up layer, the activation function layer, the  projection-down layer and a skip connection between the projection-up and projection-down layers.  The projector first projects the original $d$-dimensional channels into a larger dimension with the $1\times1 \times d \times f$
($f=2d$ in this paper) convolutional operations. Following by the non-linearity, it then projects the $f$ channels
back to the original  $d$ dimensions with the  $1\times1 \times f \times d$ convolutional operations. The output of the projector is given as 
\begin{equation}
\label{projector}
\begin{aligned}
\mathcal{F}_{PR}(\bm{\mathit{\widetilde{E}}}_L^{\tilde{x}},\bm{\mathit{\widehat{E}}}_L^{{x}})=\mathcal{F}_{agg}(\bm{\mathit{\widetilde{E}}}_L^{\tilde{x}},\bm{\mathit{\widehat{E}}}_L^{{x}}) +\phi_{down}(RELU(\phi_{up}(\mathcal{F}_{agg}(\bm{\mathit{\widetilde{E}}}_L^{\tilde{x}},\bm{\mathit{\widehat{E}}}_L^{{x}}))))
\end{aligned}
\end{equation}
where 
\begin{equation}
\label{aggregate}
\begin{aligned}
\mathcal{F}_{agg}(\bm{\mathit{\widetilde{E}}}_L^{\tilde{x}},\bm{\mathit{\widehat{E}}}_L^{{x}})=\mathcal{F}_{encoder}(\bm{\mathit{\widetilde{E}}}_L^{\tilde{x}})+\bm{\mathit{\widehat{E}}}_L^{{x}}
\end{aligned}
\end{equation}
where $\phi_{up}$ and $\phi_{down}$ represent the projection-up and projectin-down operations.

\noindent \textbf{\textbf{Model Training \& Generating}.}
As mentioned in Eq.~(\ref{gfedloss}), GRec only takes the masked positions into consideration rather than the complete sequence.
Hence, we first perform the look-up table  by retrieving the hidden vectors of the masked positions from $\mathcal{F}_{decoder}(\bm{\mathit{\widetilde{E}}}_L^{\tilde{x}},\bm{\mathit{\widehat{E}}}_L^{{x}})$, denoted by $\mathcal{F}_{decoder}^{x_\triangle} (\bm{\mathit{\widetilde{E}}}_L^{\tilde{x}},\bm{\mathit{\widehat{E}}}_L^{{x}})$. Then, we feed these  vectors into a fully-connected neural network layer which  projects them from the $d$-dimentional latent space  to the $n$-dimentional softmax space. The calculated probabilities of the masked items $x_{\triangle }$ are given as
\begin{equation}
\label{projector}
\begin{aligned}
 p(x_\triangle|\tilde{x}; \Theta)=softmax (\mathcal{F}_{decoder}^{x_\triangle} (\bm{\mathit{\widetilde{E}}}_L^{\tilde{x}},\bm{\mathit{\widehat{E}}}_L^{{x}})\bm{\mathit{W}}+\bm{\mathit{b}})
\end{aligned}
\end{equation}
where $\bm{\mathit{W}}\in \mathbb{R}^{d\times n }$ and $\bm{\mathit{b}} \in \mathbb{R}^n$ are weight matrix and the corresponding bias term. Finally,
we are able to optimize Eq.~(\ref{gfedloss}) by gradient ascent (or gradient descent on the negative of Eq.~(\ref{gfedloss})) .
Since GRec only estimates portions of items in each batch, it needs more training steps to converge compared to NextItNet-style models (which estimate the entire sequence), but much fewer steps compared to the left-to-right data augmentation based models (e.g., Caser) (which estimate only the last item).

Once the model is well trained,  we can use it for item generation. Unlike the training phase during which the encoder has to mask a certain percentage of items, GRec is able to directly compute the softmax  of the final position of the final layer 
at the inference time without performing  mask operations.

\subsubsection{Connection to Existing Models}
\label{connections}
Our work is closely related to NextItNet and the well-known bidirectional language model BERT~\cite{devlin2018bert,sun2019bert4rec}.
In this subsection, we  show the  connections of GRec to NextItNet-style and   BERT-style  models. For clarity, we omit the projection layers during discussion.

\begin{figure}[!t]
	\centering
	\subfloat[\scriptsize  GRec with one masked item ]{\includegraphics[width=0.25\textwidth]{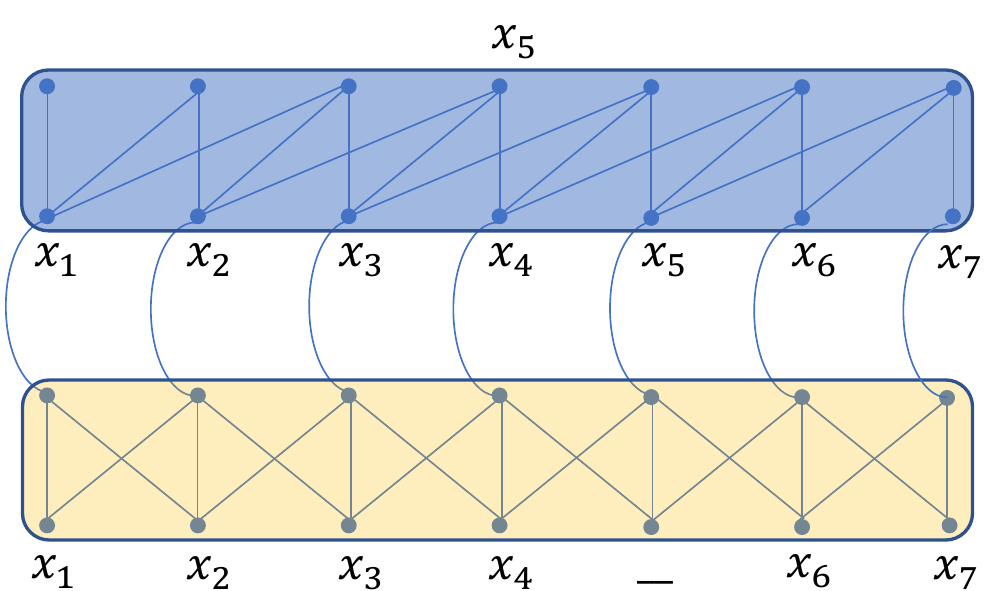}\label{one}}
	\subfloat[\scriptsize GRec with $t$ masked items. ]{\includegraphics[width=0.25\textwidth]{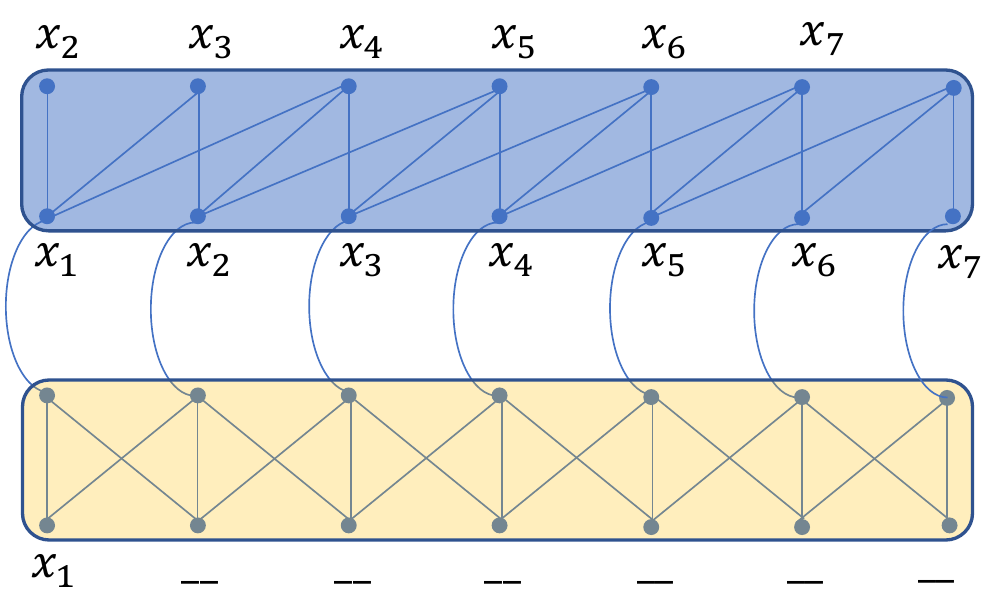}\label{all}}
	\caption{\small  GRec variants by changing the gap-filling strategy}
	\label{variant}
\end{figure}

As shown in Figure~\ref{variant} (a), when $m=1$, the encoder  masks only one position, i.e., $x_5$, from the input sequence, and correspondingly the decoder only predicts this masked token, conditioned on all other items in this sequence. 
Mathematically, we have $p(x_{\triangle }|\tilde{x})$ = $p(x_{5 }|x \backslash x_5)$. 
If we further mask the  input of the decoder with `$\_\_$', GRec reduces to a standard encoder with one softmax output, and is very similar to the well-known bidirectional language model BERT, with the only exception that  BERT applies the Transformer~\cite{vaswani2017attention} architecture while GRec uses the stacked 1D dilated CNNs. 
In this simple case, GRec reduces to a sequence-to-one model and loses its autoregressive property. In fact, The DA-based recommendation mdoels, such as Caser, IGRU4Rec and NARM, can also be seen as   sequence-to-one models which just apply different neural network infrastructures.

When $m=t-1$, all items (except the first position) in the encoder will be masked, and the decoder will predict these masked items from $x_2$ to $x_t$, as illustrated in Figure \ref{variant} (b). In this case,  the encoder of GRec becomes almost ineffective.
If we remove the encoder, GRec becomes exactly NextItNet. Note that GRec  with $m=t-1$ is very likely to perform worse than NextItNet. This is because  in such case the encoder of GRec
 introduces many additional noises, which makes the decoder much harder to be optimized.

In summary, our proposed GRec can be seen as a pseq2pseq model that jointly trains the encoder and decoder for the sequential recommendation tasks. In contrast to NextItNet-style models, GRec is able to model both the past and future contexts.  In contrast to   BERT-style models, GRec is more suitable to  the  generation task due to its  autoregressive process.  In contrast to the standard seq2seq encoder-decoder models, GRec does not have the data leakage problem when the encoder and decoder are fed into the same input sequence.

\section{Experiments}
\label{EXPERIMENTS}

As the key contribution of this work is to improve the existing left-to-right style learning algorithms for SRS, we
evaluate GRec on  real-world datasets with short-, medium- and long-range sessions,
and conduct extensive ablation studies  to answer the following research questions:

\begin{enumerate}
  \item \textbf{RQ1:} 
   Whether  the three proposed  approaches perform better than the existing left-to-right sytle models?
   Which way performs best?
  \item \textbf{RQ2:}  How does  GRec perform  with different gap-filling strategies? 
    \item \textbf{RQ3:}  What are the effects of other key modules of GRec? For example, does it benefit from the proposed projector module?
    \item \textbf{RQ4:} Is GRec a general framework or does it work well by replacing the encoder and decoder with other types of neural networks? 
\end{enumerate}

\subsection{Experimental Settings}

\subsubsection{\textbf{Datasets}} We conduct experiments on two real-world datasets with three different session lengths.  

\begin{table} 
	\centering
	\caption{\small Statistics of  the datasets.  ``M'' is short for million.}
	\small
	\label{sessionsta}
	\begin{threeparttable}				
		\begin{tabular}{@{} L{10mm}| C{15mm} |C{15mm} |C{15mm}| C{15mm}@{}}
			\toprule
			\small DATA &  \small \emph{\#actions  }&  \small \emph{\#sequences} &  \small \emph{\#items}&  \small \emph{k}\\
			\midrule
			TW10        &9,986,953 &1,048,575 & 65,997& 10\\ 	
			\midrule
			ML30      & 25,368,155 & 858,160 &18,273& 30\\ 	
			\midrule
			ML100    &25,240,741& 300,624& 18,226&100 \\

			\bottomrule
		\end{tabular}
	\end{threeparttable}
\end{table}
\begin{table*} 
	\centering
	\caption{\small Accuracy comparison. MostPop returns item lists ranked by popularity. For each measure, the best result is indicated in bold.
	}
	\small
	\label{overallresults}
	\begin{threeparttable}				
		\begin{tabular}{@{} L{15mm} | L{15mm} | C{15mm} | C{15mm}|  C{15mm}| C{15mm}| C{15mm}| C{16mm} @{}}
			\toprule
			\small DATA &  \small \emph{Models} &  \small \emph{MRR@5}&  \small \emph{MRR@20} &  \small \emph{HR@5}&  \small \emph{HR@20}&  \small \emph{NDCG@5}&  \small \emph{NDCG@20} \\
			\midrule
			
			\multirow{6}*{TW10}       &   	\emph{MostPop} & 0.0055&0.0127 & 0.0203& 0.0970&0.0091& 0.0305  \\ 
	~            & 	\emph{Caser} & 0.0780&0.0916& 0.1330& 0.2757&0.0916& 0.1317  \\ 
&\emph{GRU4Rec}  &   0.0786& 0.0926& 0.1325 & 0.2808& 0.0919 &0.1335 \\ 
			~               & \emph{NextItNet}&0.0848&0.0992 & 0.1408& 0.2931&0.0986& 0.1414  \\
			~               & \emph{NextItNet+}&0.0698&0.0844 & 0.1214& 0.2775&0.0825& 0.1218  \\ 
			~           &  	\emph{tNextItNet}   & 0.0813&0.0958 & 0.1376& 0.2896&0.0953& 0.1380  \\ 
			~           &  	\emph{GRec}& \textbf{0.0901} & \textbf{0.1046}  &\textbf{ 0.1498} &\textbf{ 0.3021}&\textbf{0.1049} & \textbf{0.1477}  \\
			\midrule
			
			\multirow{6}*{ML30}        &   	\emph{MostPop} & 0.0030&0.0058 & 0.0098& 0.0405&0.0047& 0.0132 \\  
 & 	\emph{Caser} & 0.0622&0.0739& 0.1074& 0.2323&0.0733& 0.1083 \\
	~              &\emph{GRU4Rec}  &   0.0652& 0.0788 & 0.1156 & 0.2589& 0.0776 &0.1179\\  
			~               & \emph{NextItNet}& 0.0704&0.0849 & 0.1242& 0.2756&0.0837& 0.1263  \\ 
			~               & \emph{NextItNet+}& 0.0564&0.0711 & 0.1051& 0.2609&0.0685& 0.1121  \\ 
			~           &  	\emph{tNextItNet}   & 0.0658&0.0795 & 0.1164& 0.2605&0.0782& 0.1188  \\ 
			~           &  	\emph{GRec}& \textbf{0.0742} & \textbf{0.0889}  &\textbf{ 0.1300} & \textbf{ 0.2850}&\textbf{0.0879} &\textbf{ 0.1315}\\
			\midrule
			\multirow{6}*{ML100 }        &   	\emph{MostPop} & 0.0025 & 0.0045 & 0.0076&0.0301&0.0037& 0.0099 \\ 
 & 	\emph{Caser} & 0.0492&0.0605& 0.0863& 0.2074&0.0584& 0.0922 \\
	~               & \emph{GRU4Rec} & 0.0509&0.0632& 0.0909& 0.2211&0.0608& 0.0974 \\
			~               & \emph{NextItNet} & 0.0552&0.0687& 0.1007& 0.2411&0.0664& 0.1059  \\
			~               & \emph{NextItNet+}& 0.0487& 0.0615&0.0942 & 0.2321&0.0600&0.0983  \\
			~           &  	\emph{tNextItNet}   & 0.0518&0.0642& 0.0927& 0.2239&0.0619& 0.0986  \\   
			~           &  	\emph{GRec}& \textbf{0.0588} & \textbf{0.0720}  &\textbf{ 0.1057} &\textbf{ 0.2477}&\textbf{0.0702} & \textbf{0.1101}  \\
			\bottomrule
		\end{tabular}
	\end{threeparttable}
\end{table*}

\noindent \textbf{\textbf{ ML-latest}}\footnote{\scriptsize \url{http://files.grouplens.org/datasets/movielens/}}. This  dataset 
was created on September 26, 2018 by MovieLens. Since the original dataset contains cold items, we perform a basic preprocessing by filtering out items that appear less than 20 times, similar to ~\cite{tang2018caser}.
We then generate the  interaction sequence of the same user according to the chronological order.
We split the sequence into subsequence every $k$ movies.
If the length of the subsequence is less than $k$, we pad with zero in the beginning of the sequence to reach $k$. For those with length less than $l$, we simply remove them in our experiments. In our experiments, we set $k$=30 with $l$=10 and $k$=100 with $l$=20, which results in two datasets, namely, ML30 and ML100.

\noindent \textbf{\textbf{TW10}}\footnote{\scriptsize \url{https://weishi.qq.com}}.  This  is a private dataset which 
was created on October, 2018 by the Weishi Team at Tencent Inc.. TW10 is a short video dataset, in which
the averaging playing time of each video is less than 30 seconds.
Since the cold users and items have been trimmed by the official provider, we do not need to consider the cold-start problem. Each user sequence contains 10 items at maximum.  Table~\ref{sessionsta} summarizes the statistics of evaluated datasets in this work.

\subsubsection{\textbf{Evaluation Protocols}} We randomly split all user sequences into training (80\%), validation (10\%) and testing  (10\%)  sets. We evaluate all models by three popular top-$N$ metrics, namely MRR@$N$ (Mean Reciprocal Rank),  HR@$N$ (Hit Ratio)  and NDCG@$N$ (Normalized Discounted Cumulative Gain)~\cite{yuan2018fbgd,yuan2016lambdafm,yuan2019simple}.  $N$ is set to $5$ and $20$ for comparison. The HR intuitively measures whether the 
ground truth item is on the top-$N$ list, while the NDCG \& MRR account for the hitting position by rewarding higher scores to hit at a top rank.
For each user sessions in the testing sets, we
evaluate the accuracy of the \emph{last} (i.e., next) item following~ \cite{yuan2019simple,kang2018self}.

\subsubsection{\textbf{Compared Methods}} We compare  the proposed augmentation methods with three typical sequential recommendation baselines, namely, GRU4Rec~\cite{hidasi2015session}, Caser~\cite{tang2018caser} and NextItNet~\cite{yuan2019simple}, particularly with NextItNet since GRec can be seen as an extension of NextItNet. We train Caser using the data augmentation method, and train GRU4Rec
and NextItNet based on the AR method. For fair comparisons, all methods use the cross-entropy loss function. 

\begin{figure*}
	\small
	\centering     
	\subfloat[\scriptsize  TW10 ]{\label{yahoo-alphazero}\includegraphics[width=0.3\textwidth]{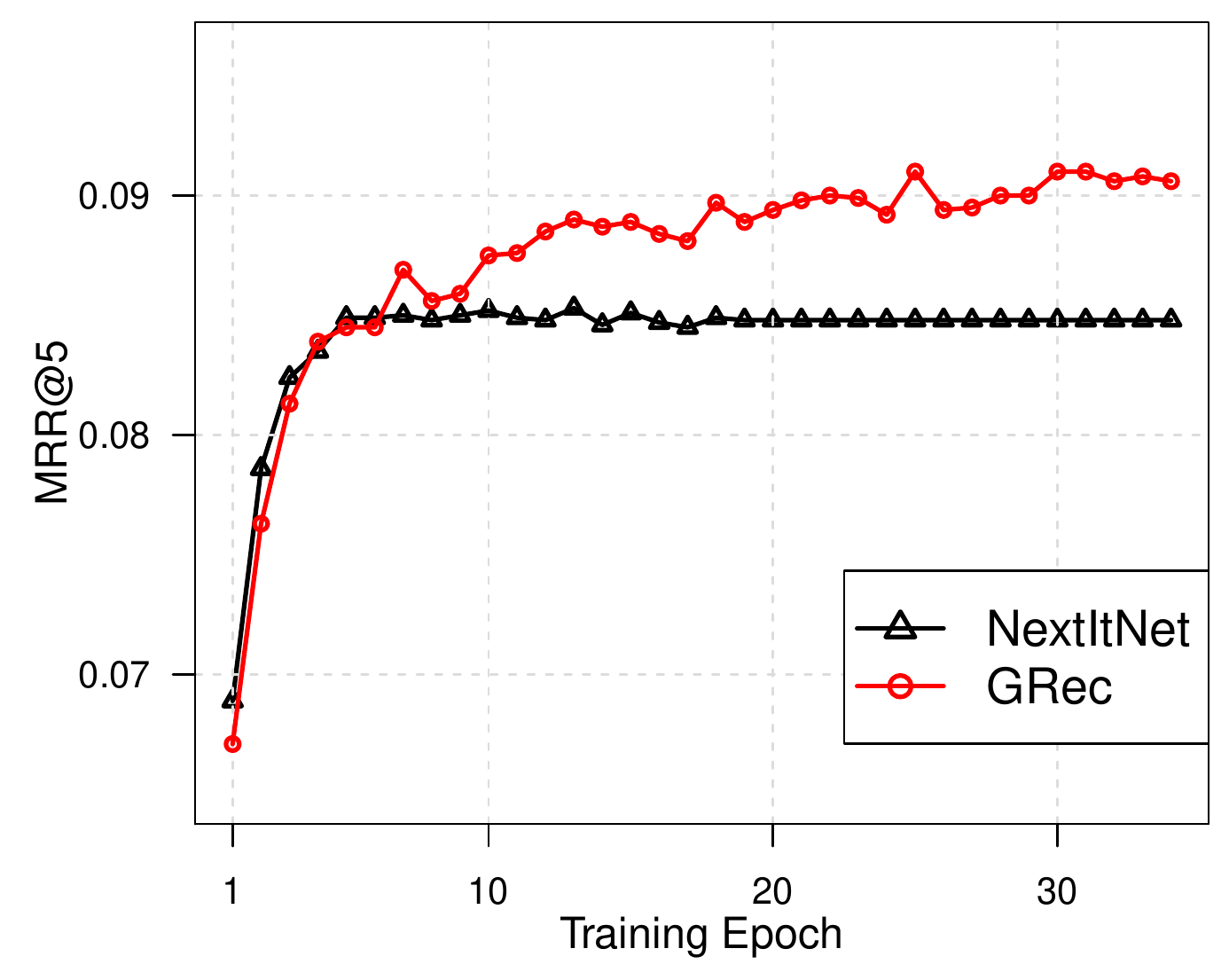}}
	\subfloat[\scriptsize  ML30]{\label{yahoo-alpha}\includegraphics[width=0.3\textwidth]{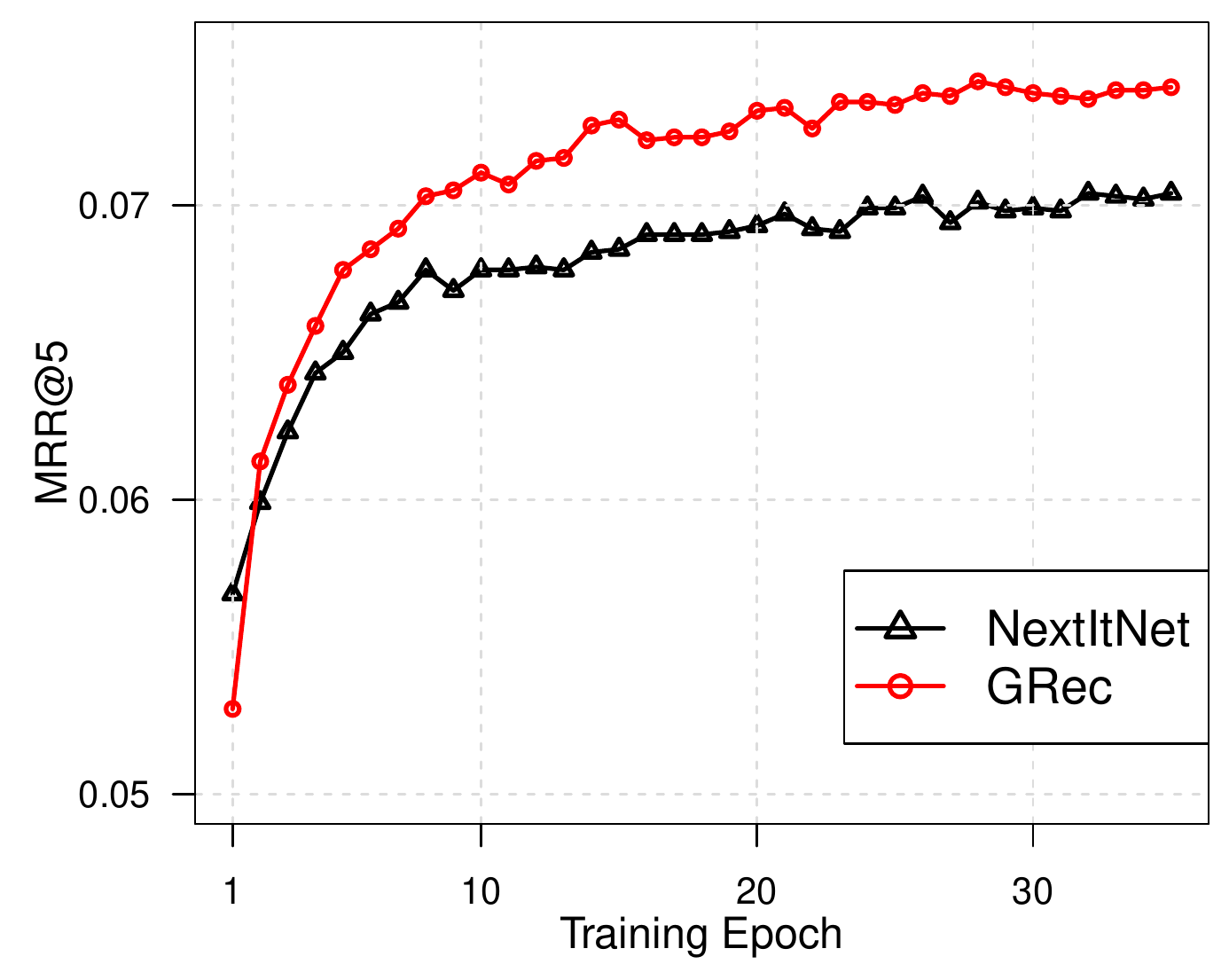}}
	\subfloat[\scriptsize  ML100]{\label{yahoo-alpha}\includegraphics[width=0.3\textwidth]{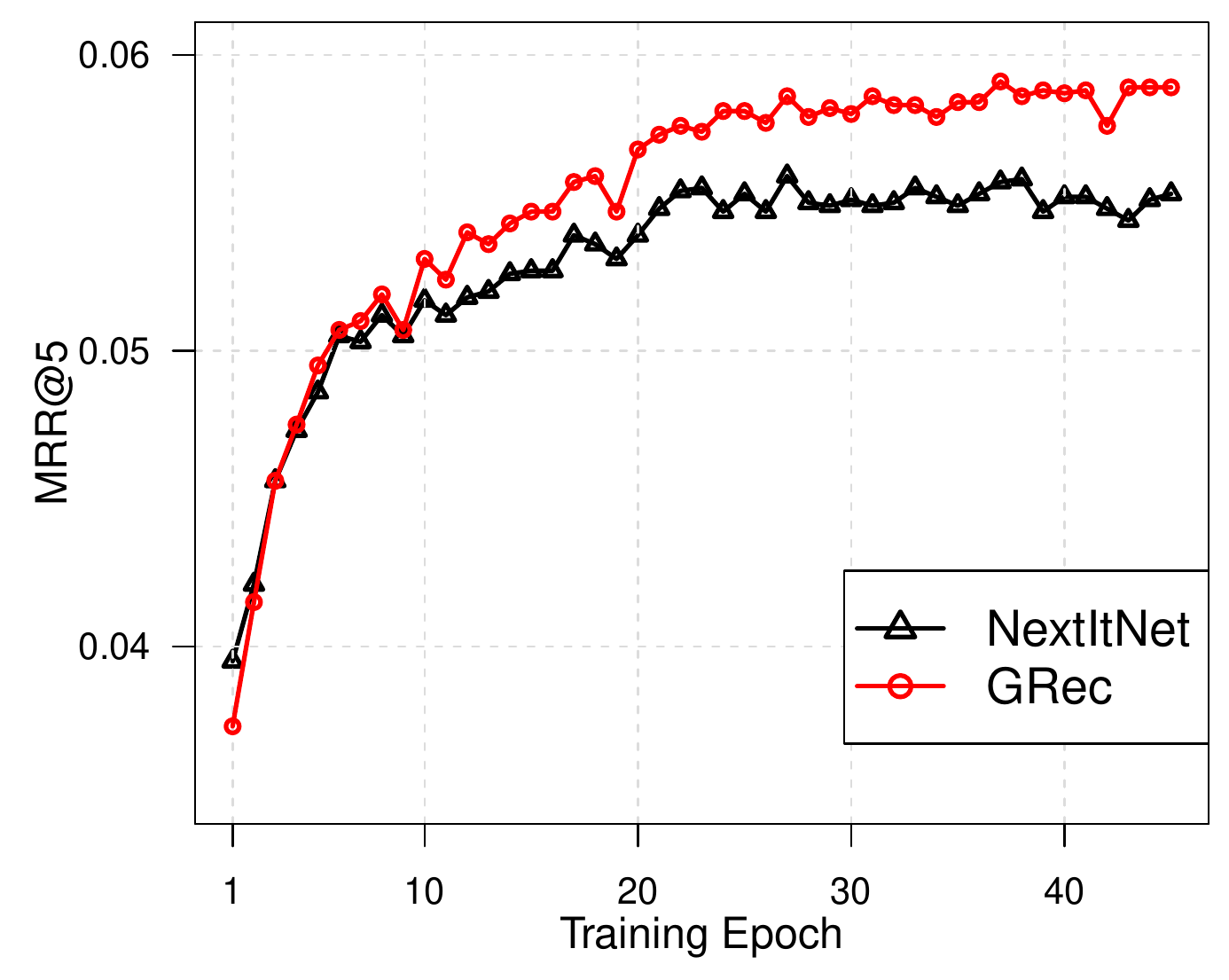}}
	\caption{\small Convergence behaviors of GRec and NextItNet. All hyper-parameters  are kept the same for the two models. 
		One training epoch in x-axis is 10000$*$128, 10000$*$256, and 3000$*$256  sequences on TW10, ML30 and ML100 respectively, where 128 and 256  are the batch size. Note we perform early stop on TW10 after 20 epoches when NextItNet fully converges  and plot the same results in the following epoches, as shown in  (a).} 
	\label{convergence1}
\end{figure*}

\begin{figure*}
	\small
	\centering     
	\subfloat[\scriptsize  TW10 ]{\label{yahoo-alphazero}\includegraphics[width=0.3\textwidth]{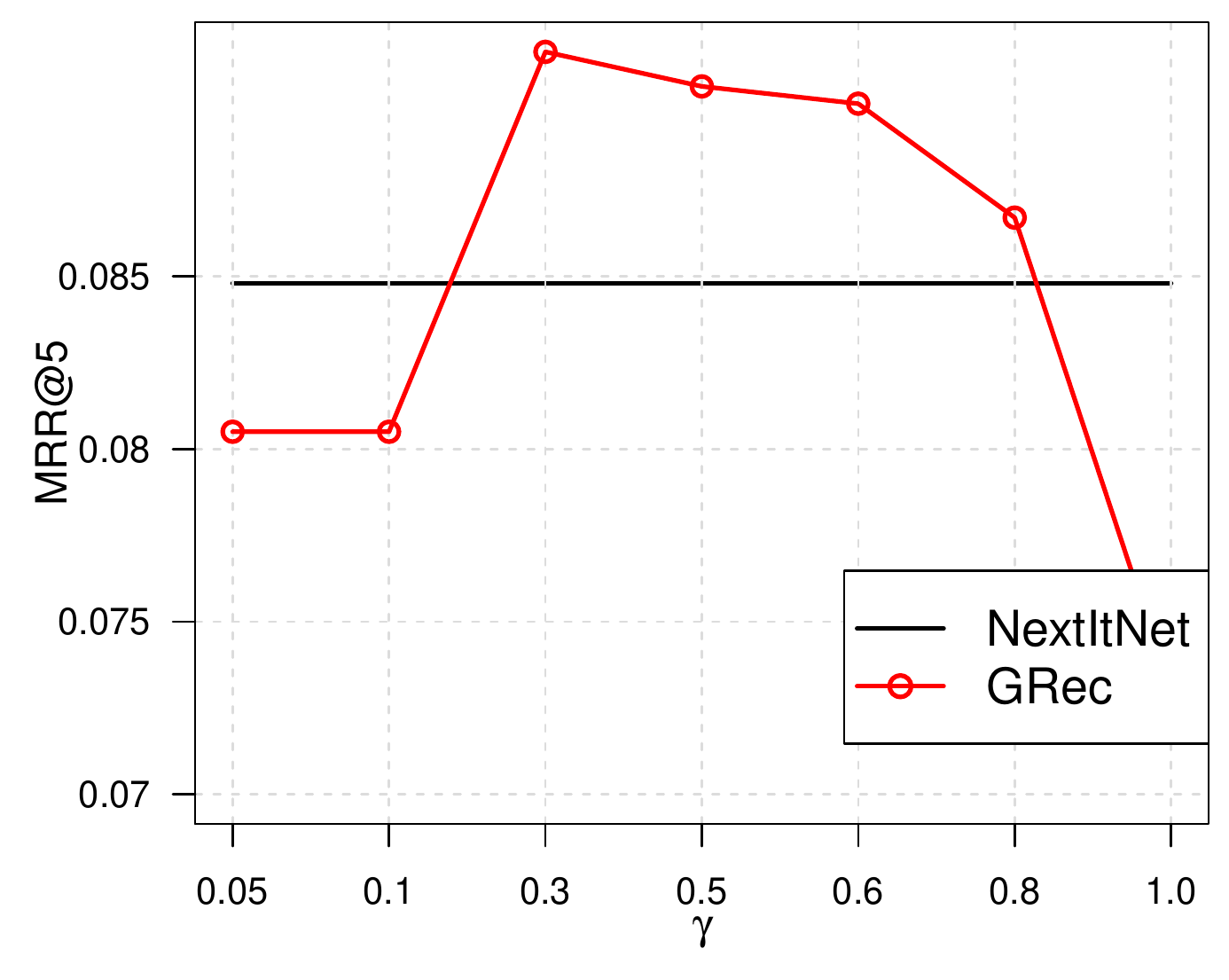}}
	\subfloat[\scriptsize  ML30]{\label{yahoo-alpha}\includegraphics[width=0.3\textwidth]{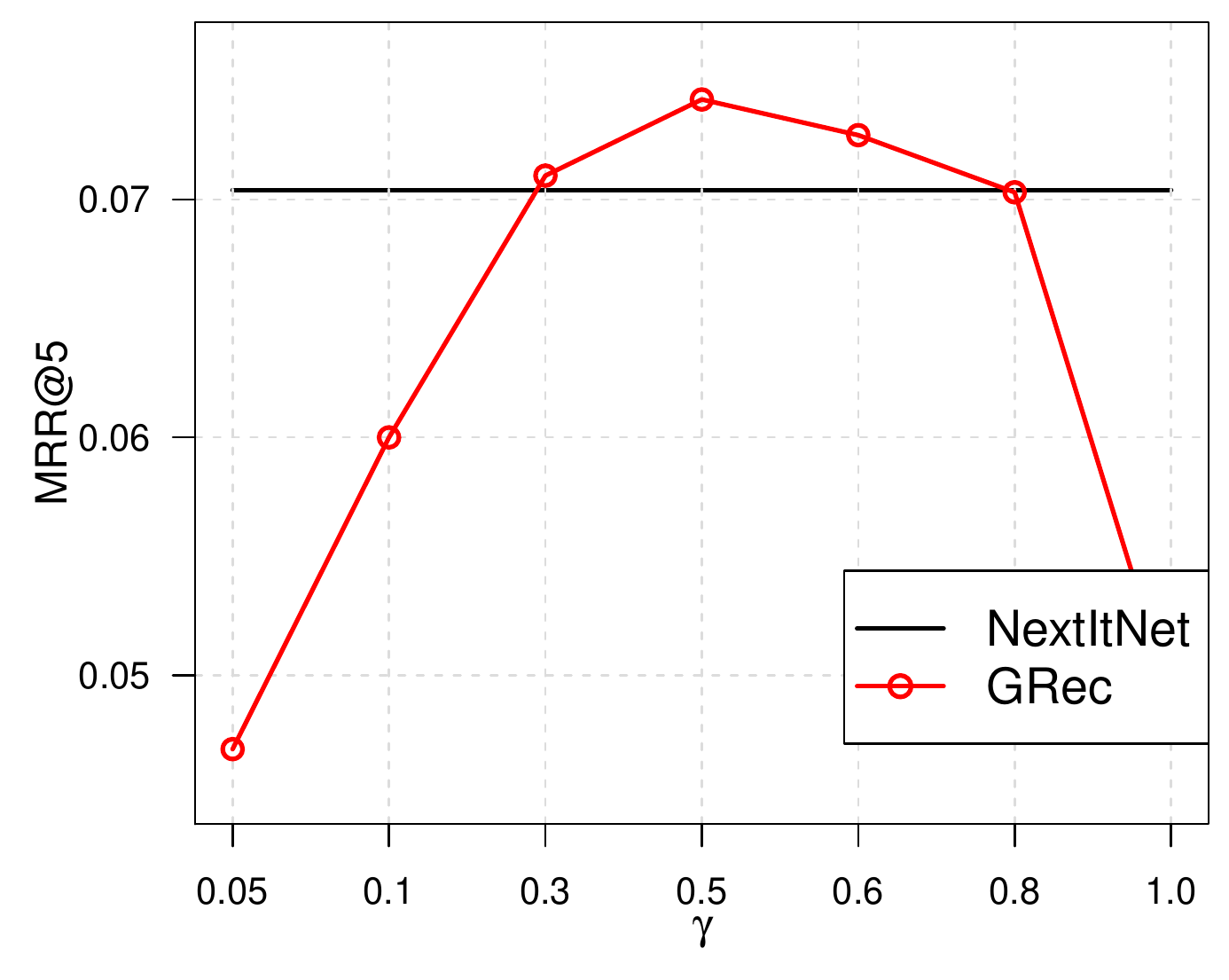}}
	\subfloat[\scriptsize  ML100]{\label{yahoo-alpha}\includegraphics[width=0.3\textwidth]{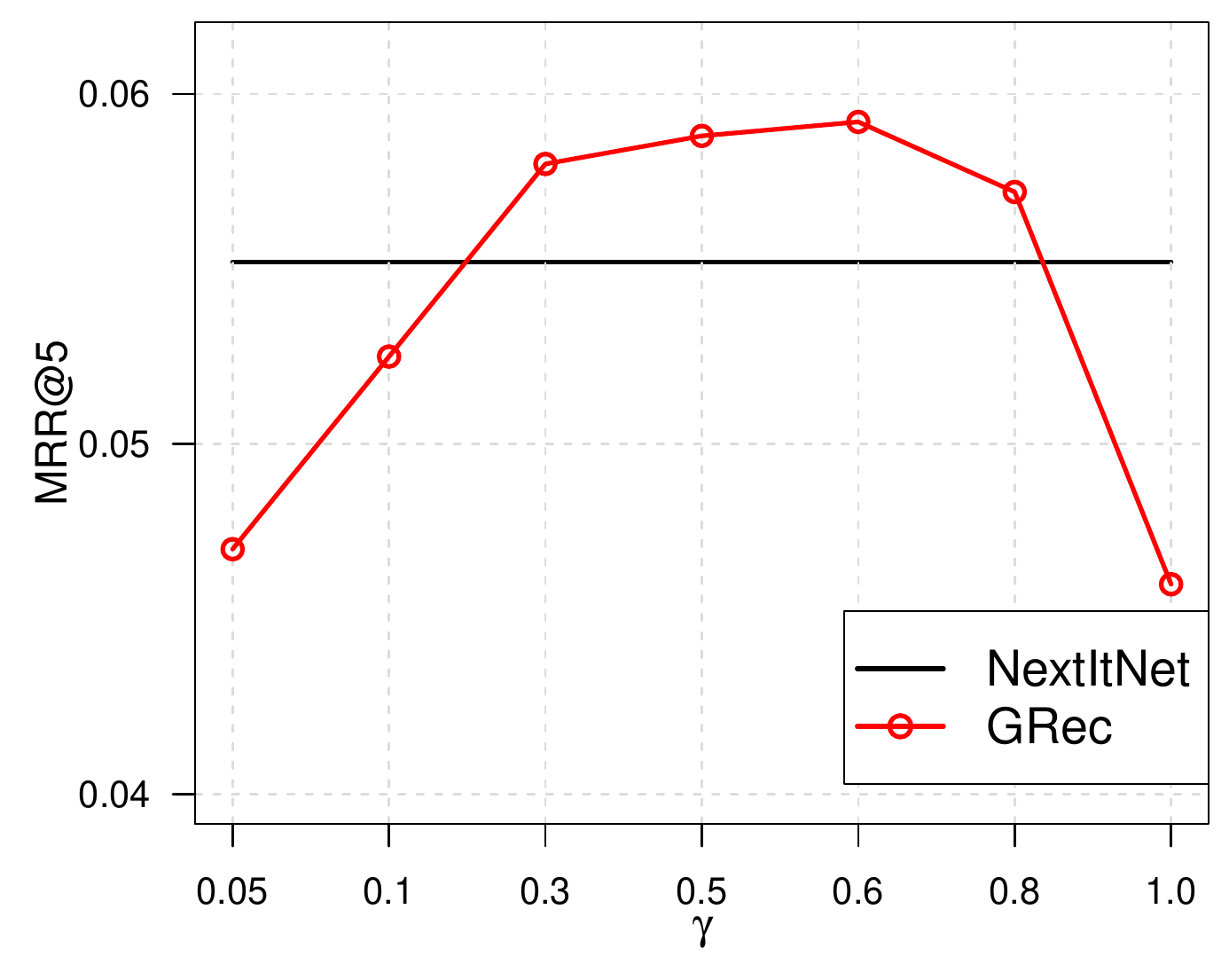}}
	\caption{\small Performance trend of GRec by tuning the percentage of masked items in the input sequence. All other hyper-parameters are kept unchanged.} 
	\label{maskper}
\end{figure*}

\subsubsection{\textbf{Implementation Details}}
 For comparison purpose, we follow  the common practice in ~\cite{kang2018self,ma2019hierarchical,rendle2014improving,wang2017irgan} by setting the embedding dimension $d$ to 64 for all models. The  hidden dimensions are set the same value as  embedding dimension $d$.  Though methods with other $d$ (e.g., $d=16, 256, 512$) yield different results, the performance trend keeps similar. 
 The learning rate is set to 0.001 in this paper.
Other  hyper-parameters of baseline methods are empirically tuned according to performance on validation sets.
NextItNet+, tNextItNets and GRec use exactly the same hyper-parameters ($q=128$ for TW10,  $q=256$ for ML30 and ML100)  as NextItNet since they can be regarded as  variants of NextItNet. 
 The dilated convolution kernels for both the encoder and decoder are set to 3.
 The dilated convolutional layers are stacked using dilation factors $\{1,2,2,4,1,2,2,4,1,2,2,4,\} $ (6 residual blocks with 12 CNN layers), $\{1,2,4,8,1,2,4,8\} $ (4 blocks with 8 CNN layers), and  $\{1,2,4,8,1,2,4,8,1,2,4,8\} $ (6 blocks with 12 CNN layers) on
  TW10, ML30 and ML100 respectively. 
 We perform sampled softmax \cite{jean2014using} on TW10 and full softmax on ML30 and ML100 for  NextItNet, NextItNet+, tNextItNets and GRec  throughout this paper.
 All models use the Adam~\cite{kingma2014adam} optimizer. All results of GRec use $\gamma=50\%$ as the gap-filling percentage without special mention.

\subsection{Performance Comparison (RQ1)}
\label{resultsumm}
Table~\ref{overallresults} presents the results of all methods on three datasets, namely, the short-range session dataset TW10, medium-range  ML30, and long-range ML100.  We first observe that  NextitNet achieves significantly better results than Caser and GRU4Rec on all three datasets.
This is consistent with the observation in~\cite{yuan2019simple}  since (1) with a fair comparison setting, the AR-based optimization method is usually more effective than the data augmentation based method  for  the sequence generating task; (2) the stacked dilated residual block architecture in NextItNet is capable of capturing more complex and longer
 sequential dependencies, while the max-pooling operations and shallow structure in Caser inevitably lose many important temporal signals and are far from optimal \cite{tang2019towards}, particularly for modeling long-range sequences.

In what follows, we focus on comparing our proposed methods  with NextItNet as they use similar neural network modules and the same hyper-parameters. First, among the three proposed augmentation methods, NextItNet+ and tNextItNet yield consistently  worse results than NextItNet, whereas GRec outperforms 
NextItNet by a large margin. 
The results of NextItNet+ and tNextItNet  indicate that  the trivial  two-way augmentation methods are not enough to guarantee better recommendation accuracy compared with the unidirectional model, although they are trained  with more data or more parameters. 
 The results are predictable since,  as we  mentioned before, the parameters learned by the right contexts in NextItNet+ may  be incompatible with those learned from the left contexts using the same convolutional filters.
 Even though tNextItNet applies two independent networks, the discrepancies during training and inference phases 
are very harmful for the recommendation accuracy.
 
 \begin{table} 
 	\centering
 	\caption{\small Impact of the projector module regarding MRR@5. NextItNetP represents NextItNet with projector. GRecN represents GRec without projector.
 	}
 	\small
 	\label{projectorablation}
 	\begin{threeparttable}				
 		\begin{tabular}{@{} L{14mm}  | C{14mm}  | C{14mm}  | C{14mm} |  C{14mm} @{}}
 			\toprule
 			\small DATA   &  \small \emph{NextItNet} &  \small \emph{NextItNetP} &  \small \emph{GRec} &  \small  \emph{GRecN} \\
 			\midrule
 			
 			{TW10}          &  0.0848& 0.0843 &0.0901&0.0880 \\ 
 			{ML30}           &  0.0704&0.0702& 0.0742& 0.0720\\
 			{ML100}           &  0.0552& 0.0558& 0.0588& 0.0577\\
 			\bottomrule
 		\end{tabular}
 	\end{threeparttable}
 \end{table}
  Second, we observe that GRec with the pseq2pseq structure signficantly exceeds NextItNet, as demonstrated in Table~\ref{overallresults}.  The results indicate that an appropriate way of modeling by using additional (i.e., future) contextual features does improve the recommendation accuracy for the unidirectional model which attends to only the past contexts. Moreover, we plot the convergence comparison of GRec and
NextItNet in Figure~\ref{convergence1}. The results show that NextItNet converges a bit faster and better than GRec in the first several epoches, but shows  poorer results than GRec after more training epoches.
The slightly slower convergence behavior is  because the loss function of GRec only considers a partial sequence whereas NextItNet loss leverages the loss of complete sequence during training  (also refer to Eq.~(\ref{compare})).  But obviously, the improved performance gains of GRec far outweigh the marginally increased training cost.

\subsection{Impact of the Gap-filling Strategies (RQ2)}

Table \ref{maskper} shows the performance change of GRec with different gap-filling percentages. We fix all other hyper-parameters by tuning $\gamma$. As clearly shown, too large or too small $\gamma$ typically achieves suboptimal performance. The highest recommendation accuracy is obtained when  $\gamma$ is  between $30\%$ to $50\%$. The is because masking too much percentage of items  in the user session is very likely to (1)
 discard important future contexts; (2)  introduce noises due to the masked tokens; and (3) 
bring more discrepancies between training and inference phases, as explained in Section~\ref{connections}. 
E.g., when  $\gamma=1.0$, no future contexts are leveraged, and the encoder of GRec becomes a neural network with only noises. In this case, GRec performs even worse than the standard NextItNet. 
On the other side, GRec will lose its autoregressive advantage when $\gamma$ is smaller, and  becomes an simple encoder network or a sequence-to-one model when only one item is masked. With this setting, the sequential and recurrent patterns will not be captured any more. Hence, there is a clear trade-off for GRec between taking advantage of future contexts and making use of the autoregressive property.

\begin{table} 
	\centering
	\caption{\small GRec vs. its encoder regarding MRR@5. $\gamma$ is set to 0.5 for both GRec and its encoder.
	}
	\small
	\label{encodercomp}
	\begin{threeparttable}				
		\begin{tabular}{@{} L{14mm}  | C{14mm}  | C{14mm}  | C{14mm}  @{}}
			\toprule
			\small DATA   &  \small TW10 &  \small ML30&  \small ML100 \\
			\midrule
			
			\emph{GRec}          &  0.0901& 0.0742 &0.0588 \\ 
			\emph{Encoder}           &  0.0808&0.0592& 0.0489\\
			\bottomrule
		\end{tabular}
	\end{threeparttable}
\end{table}

\begin{table} 
	\centering
	\caption{\small GRec vs. NextItNet with $d=512$ regarding MRR@5.
	}
	\small
	\label{larged}
	\begin{threeparttable}				
		\begin{tabular}{@{} L{14mm}  | C{14mm}  | C{14mm}  | C{14mm}  @{}}
			\toprule
			\small DATA   &  \small TW10 &  \small ML30&  \small ML100 \\
			\midrule
			
			\emph{NextItNet}          &  0.111& 0.105 &0.093 \\ 
			\emph{GRec}           &  0.117&0.114& 0.103\\
			\bottomrule
		\end{tabular}
	\end{threeparttable}
\end{table}
\subsection{Ablation Studies (RQ3)}

In this subsection, we first investigate the effectiveness of the projector module. One may argue that the improved performance of GRec relative to NextItNet may come from the additional projector  module.  To clear up the uncertainty, we perform a fair ablation study by removing the projector for GRec as well as injecting it for NextItNet. We have the following observations according to Table~\ref{projectorablation}: (1) The projector indeed helps GRec achieve better performance by comparing 
GRec \& GRecN; (2) NextItNet is inferior to GRec even with the projector by comparing GRec \& NextItNetP; (3) GRec still exceeds NextItNet even without the projector  by comparing GRecN \& NextItNet.

Second, we also report results of GRec with only the  encoder network since the encoder itself is also able to leverage two directional contexts. To do so, we remove the decoder of GRec and place the softmax layer on the encoder during training. 
At the generating phase, we just need to replace the last item by "$\_\_$", and retrieve the top-$N$ scored items for comparison. With this special case,
GRec reduces to a BERT-like bidirectional encoder model.
We report the MRR@5 in Table~\ref{encodercomp}. As can be seen, GRec largely exceeds its encoder on all three datasets. The findings confirm our previous analysis since the bidirectional enocder network is not autoregressive and  fail to explicitly model the sequential patterns of previous interactions over time. In addition, some of the left contexts are missing because of the gap-filling mechanism, which also results in unsatisfied performance.

In addition, we also report NextItNet \& GRec with a very large $d$  in Table~\ref{larged} (along with much longer training time  and larger memory consumption). As shown,  GRec obtains significant improvements relative to NextItNet, similar to the observations in Table~\ref{overallresults}.

\begin{table} 
	\centering
	\caption{\small Recurrent variants of GRec regarding MRR@5.. 
	}
	\small
	\label{variants}
	\begin{threeparttable}				
		\begin{tabular}{@{} L{14mm}  | C{14mm}  | C{14mm}  | C{14mm} |  C{14mm} @{}}
			\toprule
			\small DATA   &  \small \emph{ReCd} &  \small \emph{NextItNet} &  \small \emph{CeRd}&  \small  \emph{GRU}\\
			\midrule
			
			{TW10}          &  0.0879& 0.0843 &0.0876&0.0786 \\ 
			{ML30}           &  0.0728&0.0704& 0.0712& 0.0652\\
			{ML100}           &  0.0582& 0.0552& 0.0571& 0.0509\\
			\bottomrule
		\end{tabular}
	\end{threeparttable}
\end{table}

\begin{figure}[!t]
	\centering
	\subfloat[\scriptsize ReCd: BiRNN+ causal CNN. ]{\includegraphics[width=0.24\textwidth]{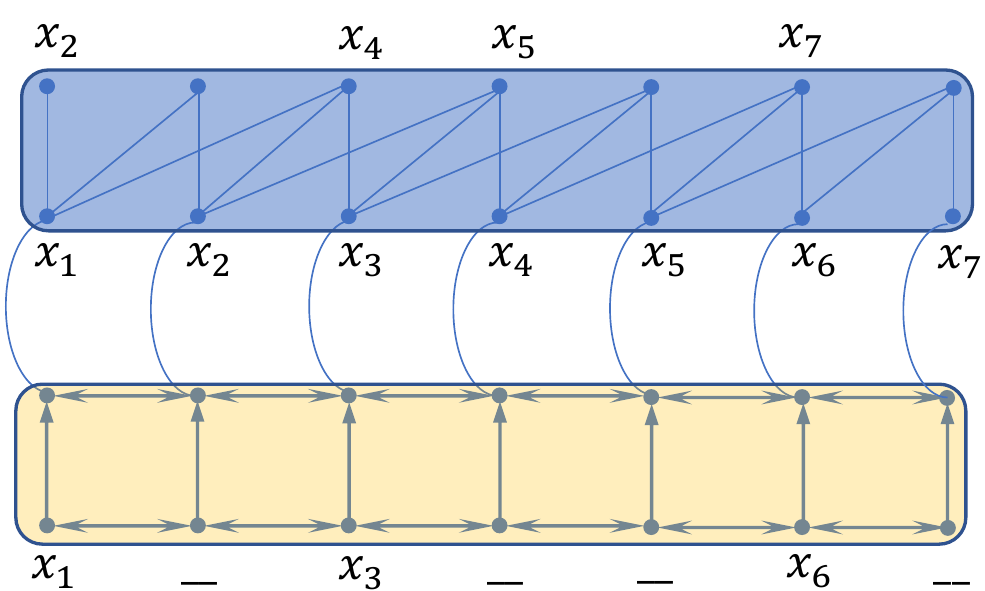}\label{one}}
	\subfloat[\scriptsize CeRd: non-causal CNN + RNN. ]{\includegraphics[width=0.24\textwidth]{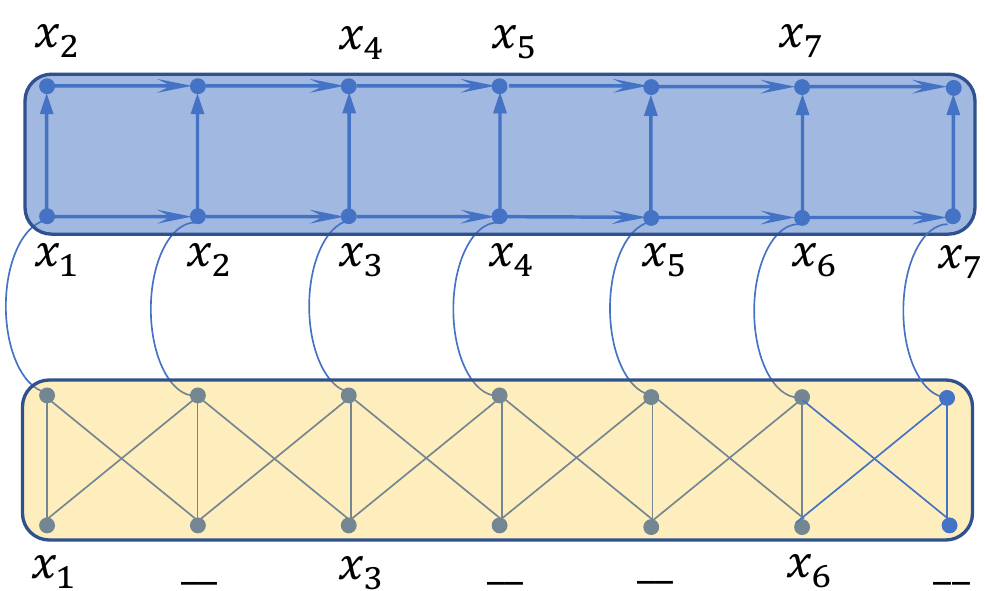}\label{all}}
	\caption{\small  GRec variants with recurrent encoder or decoder.}
	\label{ENDEvariants}
\end{figure}

\subsection{GRec Variants  (RQ4)}
Since GRec is a general ED framework, one can simply replace the original dilated convolutional neural network with other types of neural networks, such as RNN. For the sake of completeness, we demonstrate two GRec variants in 
 Figure~\ref{ENDEvariants}. ReCd represents GRec with \underline{r}ecurrent \underline{e}ncoder network (Bi-GRU) and \underline{c}onvolutional \underline{d}ecoder network, while CeRd represents GRec with \underline{c}onvolutional \underline{e}ncoder network and \underline{r}ecurrent \underline{d}ecoder network (GRU).
We report results on Table~\ref{variants} and make two observations: (1) GRec still exceeds NextItNet even it utilizes Bi-GRU as encoder by comparing ReCd \& NextItNet; (2) GRec outperforms  GRU when it utilizes the typical GRU as decoder by comparing CeRd \& GRU; (3) in general, the GRec framework using stacks of convolutional  blocks for both its encoder and decoder performs better than its variants using RNN for either encoder or decoder.
The above observations further verify the generality and flexibility of GRec for processing future contexts.

\section{Conclusion}
In this paper, we perform studies on how to incorporate future contexts for the typical left-to-right style  learning algorithms in the task of SRS.  The motivation is that the architectures of autoregressive-based sequential recommendation models  fail to model the past and future contexts simultaneously. 
To maintain the autoregressive property as well as utilize two directional contexts,  we present GRec, a novel pseq2pseq encoder-decoder neural network recommendation framework with gap-filling based optimization objective. 
GRec is general and flexible, which jointly trains the encoder and decoder on the same user action sequence without causing the data leakage issue.
Through ablations and   controlled experiments, we demonstrate that GRec is more powerful than the traditional unidirectional models. For future work, we are interested in studying whether the right contexts or GRec can improve the recommendation diversity for SRS.

\section*{Acknowledgement}
This work is supported by the National Natural Science Foundation of China (61972372, U19A2079).
\scriptsize
\bibliographystyle{ACM-Reference-Format}

\bibliography{bibliography}

\end{document}